\documentclass[english,prx,floatfix,longbibliography,floatfix,superscriptaddress,twocolumn,showpacs,amsmath,amssymb,reprint,latin9]{revtex4-2}

\usepackage[T1]{fontenc}
\usepackage[latin9]{inputenc}
\usepackage{color}
\usepackage{amsmath}
\usepackage{amssymb}
\usepackage{graphicx}

\definecolor{asparagus}{rgb}{0.53, 0.66, 0.42}
\definecolor{red}{rgb}{1,0,0}
\definecolor{dgreen}{rgb}{0,0.5,0}

\newcommand{\bk}{{\bf{k}}}

\makeatletter

\newcommand{\lyxmathsym}[1]{\ifmmode\begingroup\def\b@ld{bold}
  \text{\ifx\math@version\b@ld\bfseries\fi#1}\endgroup\else#1\fi}

\providecommand{\tabularnewline}{\\}

\PassOptionsToPackage{caption=false}{subfig} 
\usepackage{hyperref}
\hypersetup{
breaklinks=true,
colorlinks=true,
citecolor=blue,
linkcolor=blue,
filecolor=blue,
urlcolor=blue
}
\IfFileExists{lmodern.sty}{\usepackage{lmodern}}{}
\makeatother

\usepackage{babel}

\begin{document}
\title{Floquet engineering multi-channel Kondo physics}

\author{Victor  L. Quito}
\email{vquito@ifsc.usp.br}
\affiliation{Department of Physics and Astronomy, Iowa State University, Ames,
Iowa 50011, USA}
\author{R. Flint}
\affiliation{Department of Physics and Astronomy, Iowa State University, Ames,
Iowa 50011, USA}
\date{\today}
\begin{abstract}
Floquet engineering is a powerful technique using periodic potentials, typically laser light, to drive materials into regimes inaccessible in equilibrium.  Here, we show that Kondo models can be driven to multi-channel degenerate points, even when the starting model is single-channel. These emergent channels are differentiated by symmetry, and their strength and number can be controlled by changing the light polarization, frequency and amplitude.  Unpolarized light, constructed by polarization averaging, is particularly useful to induce three and four channel degeneracies.  Multi-channel Kondo models host a wide variety of exotic phenomena, including non-Abelian anyons in impurity models and composite pair superconductivity in lattice models. We demonstrate our findings on both a simple square lattice toy model and a more realistic spin-orbit coupled model for $J=5/2$ Ce ions in a tetragonal environment, as relevant for the Ce 115 materials, and show that the transition temperature for composite pair superconductivity can be dynamically enhanced.
\end{abstract}
\maketitle

\section{Introduction\label{sec:Intro}}

Heavy fermion materials are prototypical correlated electron systems that host a wide range of phenomena, including topological phases~\cite{Dzero2016,chang2017,lai2018}, unconventional superconductivity~\cite{stewart17}, and quantum criticality~\cite{coleman05,gegenwart08}. 
This physics is already present in the single-channel Kondo model~\cite{doniach77}, and even richer physics is possible with multi-channel interactions, where multiple, symmetry distinct flavors of conduction electrons screen the same local moment~\cite{CoxZawadowski1998}.  Multi-channel Kondo impurities are believed to realize non-Abelian anyons, including Majorana~\cite{EmeryKivelson_PRB_1992} and Fibonacci anyons~\cite{Lopes_PRB_2020,Komijani_PRB_2020} in the two- and three-channel cases, respectively.  
The multi-channel Kondo lattice likely includes non-Fermi liquid phases~\cite{jarrell96,cox96},  superconductivity~\cite{cox87,Coleman_PRB_1999}, channel symmetry breaking order~\cite{schauerte05,hoshino2011,ChandraFlint2013} and even more exotic physics~\cite{patri2020}.  
Unfortunately, channel degeneracies are rare in nature. Although some two-channel degeneracies occur in non-Kramers materials~\cite{CoxZawadowski1998,sakai11}, and multi-channel Kondo physics has been realized in quantum dots \cite{Goldhaber-Gordon_Nature_1998,Potok_Nature_2007,Pierre_Science_2018}, higher channel degeneracy has not yet been found in materials, and there is currently no way to continuously tune through channel degenerate points to experimentally study the full spectrum of multi-channel Kondo lattice physics.
In this paper, we argue that Floquet engineering allows versatile, continuous tuning of multi-channel Kondo physics in both the impurity and lattice cases and that the light polarization can control the number of Kondo channels. 

Floquet engineering manipulates many-body systems by driving them with periodic light so as to modify the underlying interactions and realize diverse phenomena not easily accessible in equilibrium~\cite{Oka_review_2019}.  It has been applied to tune phases from superconductors~\cite{Demler_PRB_2016,Benito_PhysE_2015,You_PRB_2017,TakasanKawakami_PRB_2017,Takasan_PRB_2017,IshiiMurata_PRD_2018,Eckstein_PRB_2018} to topological insulators~\cite{Lindner2011,Wang2013Science,Platero_PRL_2013,Neupert_PRL_2014,FietePRB2018,RudnerLindner_NatRev_2020} and time-crystals~\cite{Krzysztof_PRA_2015,Khemani2016,else2016}.  Its application to correlated materials is particularly intriguing.  In frustrated magnets, Floquet engineering can theoretically tune the underlying exchange interactions~\cite{Mentink_NatPhys_2015,SatoOka_2016_PRL,Mentink_2017,Ishihara_PRB_2018,Refael_PRB_2019,Owerre_2019,Losada_PRB_2019,Millis_PRB_2019,Refael_PRB_2019,Mentink_2019_SciPost,Fulga2019,QuitoFlintPRL2021} and induce chiral fields~\cite{ClaassenNatComm2017} by tuning the frequency, amplitude and polarization of the light.
Previous applications of Floquet to Kondo systems include tuning topological Kondo insulators~\cite{TakasanKawakami_PRB_2017}, inducing eta pairing~\cite{YunokiMiyakoshi_PRB_2020} and driving dynamical phase transitions in the quarter-filled single-channel case~\cite{fauseweh_PRB_2020}. 
Most interestingly, a quantum dot Kondo impurity in a time-periodic chemical potential can be driven through an effective two-channel critical point~\cite{eckstein2017arxiv}, where numerical simulations showed that, during the pre-thermalization regime, the driven one-channel model generically reproduces equilibrium (non-degenerate) two-channel Kondo physics. At the critical point, the relaxation time diverges and the physics is captured by a quantum quench to the two-channel critical point. Here, we generalize this simple one-dimensional case to explore all possibilities available for a two-dimensional single-channel Kondo model (impurity or lattice) driven with periodic light of different polarizations. This approach can generate \emph{all} channels available for a given lattice, and allows tuning through multi-channel degenerate points.  

\begin{figure*}
\includegraphics[width=2\columnwidth]{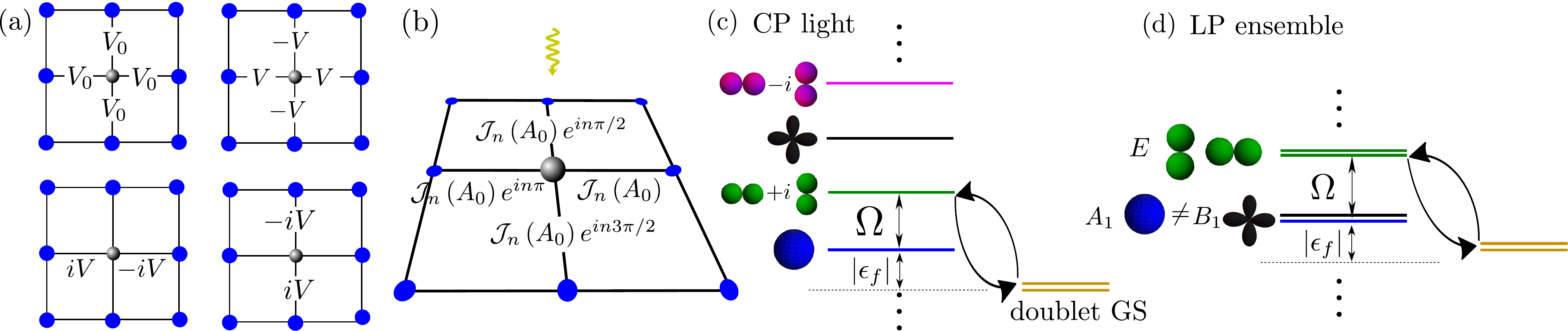}

\caption{The Floquet-Kondo model and its emergent channels. (a) We consider the equilibrium hybridization of the $f$-electron (center site) with the
conduction electrons on neighboring (blue) sites to be $s$-wave, with
a real hybridization $V_{0}$ (top left). The Floquet engineering
dresses the hybridization into d-wave (top right), and $p_{x}$ and
$p_{y}$ forms (bottom row) symmetries. (b) Photons transfer angular momentum to the electrons, leading the hybridization to acquire phases and modulations that
depend on an integer $n$ labeling the different Floquet sectors. These example phases are for the circularly polarized light case, with fluence $A_0$ and where $\mathcal{J}_{n}$ are Bessel functions. (c)-(d) The effective
low-energy Floquet-Kondo model contains different emergent channels
coming from the dressed hybridizations, mimicking models with coexisting
$s$, $p$, and $d$ orbitals. The ground state doublet is a Kramers doublet and the Floquet field generates infinite copies of the initial excited singlet with different symmetries. In (c), we show the case of circularly polarized (CP) light and in (d) an ensemble average of linearly polarized (LP) light. The cartoon orbitals show the symmetry of the hybridization that excites electrons into those excited states. \label{fig:Floquet_Kondo_schem}}
\end{figure*}

The principle is illustrated by coupling the Anderson model, precursor of the single-channel Kondo model, to periodic light, shown in Fig.~\ref{fig:Floquet_Kondo_schem}. The time-periodicity implies an infinite number of Floquet sectors in frequency space, labeled by integers, $m$.  $m$-photon-dressed hybridizations mediate virtual fluctuations from the ground state doublet to the excited singlet in the $m$-th Floquet sector, and the photon angular momentum contributes phases that change the hybridization symmetry, as seen in Fig.~\ref{fig:Floquet_Kondo_schem}.  In this work, we show that the Floquet fields generically generate all possible channels allowed by the combination of the lattice and polarization symmetries. The one-dimensional problem considered in Ref.~\cite{eckstein2017arxiv} is a particular example, allowing $s$ and $p$ wave hybridizations. The emergent Kondo couplings can be resonantly enhanced when the photon energy is near a particular fraction of the energy difference between the equilibrium ground and excited states. These resonances can selectively enhance or suppress channels and be used to tune to degenerate points.

We illustrate this physics with two models with %
$C_{4v}$ symmetry in equilibrium: a toy model on the square lattice, and a spin-orbit-coupled generalization to $J=5/2$ Ce ions in tetragonal symmetry. We also show an exact mapping between the light driven Kondo impurity and a driven quantum dot.  As polarization generically breaks lattice symmetries, we specifically consider both circular polarization, which preserves rotational symmetries, and polarization averages~\cite{LehnerPRA1996,QuitoFlintPRL2021}, which can restore the full $C_{4v}$ symmetry. The emergent Kondo couplings can therefore realize up to four channel degeneracies.
In the lattice, the multi-channel Kondo model may support composite pair superconductivity, and $d_{x^2-y^2}$ composite pairing has been proposed to cooperate with magnetic pairing in the Ce$M$In$_5$ ($M$=Co,Rh,Ir) heavy fermion materials to give robust superconductivity~\cite{Petrovic_2001,Park_Nature_2006,Flint_NatPhys_2008,Singh_2015,Prozorov_PRL_2015,Flint_NatPhys_2008}. Our driven model supports both even and odd frequency superconductivity with $d$ and $p$-wave symmetries, respectively, and we show that the $d_{x^2-y^2}$ superconducting transition temperatures can be substantially enhanced by Floquet engineering.

The remainder of this paper is organized as follows. In Section~\ref{sec:Anderson_model}, we introduce a simple Anderson model coupled to time-periodic potentials and show how to apply the Floquet theory. In Section~\ref{sec:polarization}, we consider a generic polarization of light and how it affects the dressed Wannier functions. In Section~\ref{sec:Kondo_couplings}, we perform a Schrieffer-Wolff transformation of the effective Floquet-Anderson model leading to new emergent channels. The results for the intensity and number of channels of our toy model are shown in Section~\ref{sec:toy_model_results}.  In Section~\ref{sec:quantum_dots}, we connect our results to quantum dot realizations, while in Section~\ref{sec:Ce_model} we consider a realistic model for Ce ions. The composite order of the effective Kondo lattice model is treated in Section~\ref{sec:Composite_order}. In Section~\ref{sec:exp}, we discuss how to implement our proposal experimentally, and, finally, in Section~\ref{sec:Conclusions}, we summarize our conclusions and point to future directions.

\section{Floquet-Anderson model\label{sec:Anderson_model}}

We begin with the equilibrium Anderson model, where strongly interacting $f$-electrons hybridize with non-interacting conduction electrons, $c$.  As we are interested in both impurity and lattice models, we label the $f$ sites by $i$, fixing $i = 0$ for the impurity case.
\begin{align}\label{eq:anderson}
H& =\sum_{\boldsymbol{k},\sigma}\epsilon_{\boldsymbol{k}}c_{\boldsymbol{k}\sigma}^{\dagger}c_{\boldsymbol{k}\sigma}+V_{0}\sum_{i,\sigma}\left[f_{i\sigma}^{\dagger}\psi_{i\sigma}+\text{h.c.}\right]+\nonumber \\ & -\left|\epsilon_{f}\right|\sum_{i}f_{i\sigma}^{\dagger}f_{i\sigma}+U\sum_{i}\hat{n}_{fi\uparrow}\hat{n}_{fi\downarrow}
\end{align}
Here, $\sigma$ labels the spin, $V_0$ is the bare hybridization, $|\epsilon_f|$ is the $f$-electron chemical potential and $U$ is the $f$-electron Hubbard interaction, which we will take to be infinite to %entirely prevent
block valence fluctuations into doubly occupied states.  We consider $c$- and $f$-electrons to occupy different orbitals on the same sites, such that the state that actually hybridizes with the $f$-electron is a Wannier state constructed from conduction electrons at neighboring sites,
\begin{equation}
\psi_{i\sigma}=\sum_{j}\phi_{ij}^{\sigma\sigma^{\prime}}c_{j\sigma^{\prime}},
\end{equation}
with $\phi_{ij}^{\sigma\sigma^{\prime}}=\left\langle c_{j\sigma^{\prime}}\left|\psi_{i\sigma}\right.\right\rangle$. In the toy model, we take $\phi_{ij}^{\sigma\sigma^{\prime}}=\delta_{\sigma\sigma^{\prime}}$ if $i,j$ are nearest neighbors, and zero otherwise, $\psi_{j\sigma}=\sum_{\boldsymbol{\delta}}c_{\boldsymbol{j+\delta}\sigma}$. In the Kondo limit, $\left|\epsilon_{f}\right|\gg V_0$, a Schrieffer-Wolf transformation leads to the single-channel Kondo interaction~\cite{Schrieffer1966},
\begin{equation}
H_{K}=J_{0}\sum_{\boldsymbol{k},\boldsymbol{k}^{\prime},i}e^{i\left(\boldsymbol{k}^{\prime}-\boldsymbol{k}\right)\cdot\boldsymbol{R}_{i}}\beta_{\boldsymbol{k}}\beta_{\boldsymbol{k}^{\prime}}\left(c_{\boldsymbol{k}\sigma}^{\dagger}\boldsymbol{\sigma}_{\sigma\sigma^{\prime}}c_{\boldsymbol{k}^{\prime}\sigma^{\prime}}\right)\cdot\boldsymbol{S}_{i}
\end{equation}
where $J_0=2V_{0}^{2}/\left|\epsilon_{f}\right|$ and $\beta_{\boldsymbol{k}}=\cos k_{x}+\cos k_{y}$, accounting for the extended s-wave hybridization~\cite{WeberVojta_PRB_2008}. The full Kondo model also includes the conduction electron kinetic energy from Eq.~\eqref{eq:anderson}. The Ce model generalization is similar, but with different symmetries, and is further discussed in Section~\ref{sec:Ce_model}.

Now we couple this model to a periodic 
electromagnetic field. We assume that the field propagates along the $z$ direction, perpendicular to a two-dimensional sample.
In three-dimensional systems, the propagation direction can be chosen perpendicular to crystallographic planes such that only the hybridization in those planes will be changed~\footnote{The Floquet field can, therefore, partially tune the dimensionality.}.
The field modifies the in-plane overlaps: both the conduction electron hoppings that enter into $\epsilon_{\bf k}$ and the Wannier function overlaps $\phi_{ij}$, which are changed according to the Peierls substitution~\cite{AokiPRB2016}, leading the states to acquire an explicit time dependence,
\begin{align}
\psi_{i,\sigma}\left(t\right)=\sum_{j,\sigma^{\prime}}\phi_{ij}^{\sigma\sigma^{\prime}}e^{i\int_{\boldsymbol{R}_{i}}^{\boldsymbol{R}_{j}}\boldsymbol{A}\left(t\right)\cdot d\boldsymbol{l}}c_{_{j},\sigma^{\prime}}\label{eq:cond_Floquet}.
\end{align}
$\boldsymbol{A}\left(t\right)$ is the vector potential associated with an electric field varying with frequency $\Omega$, $\boldsymbol{A}\left(t\right)=\frac{i}{\Omega}\boldsymbol{E}\left(t\right)$, which has a period $T=2\pi/\Omega$.  

The periodic nature allows us to use Floquet theory to determine the dynamics in the transient regime before heating effects become important. Generically, any time-periodic problem can be Fourier-transformed to frequency space, with discrete integer labels $m$ that index the \emph{Floquet sectors}. The time-dependent phase of Eq.~(\ref{eq:cond_Floquet}) dresses the bare hybridizations and hoppings between Floquet sectors differing by $n$ with the factor,
\begin{align}
g_{n}\left(\boldsymbol{\delta}_{ij}\right) & =\frac{1}{2\pi}\int_{0}^{2\pi}d\theta e^{-in\theta}e^{i\boldsymbol{A}(\theta)\cdot\boldsymbol{\delta}_{ij}},\label{eq:gn_Floquet}
\end{align}
where $\theta = \Omega t$ and $\boldsymbol{\delta}_{ij}=\boldsymbol{R}_{j}-\boldsymbol{R}_{i}$ connects the two neighboring sites. The Wannier functions now carry a Floquet index $n$, $\psi_{j,\sigma}^{\left(n\right)}$,
\begin{equation}
\psi_{i,\sigma}^{\left(n\right)}=\sum_{j,\sigma^{\prime}}\phi_{ij}^{\sigma\sigma^{\prime}}g_{n}\left(\boldsymbol{\delta}_{ij}\right)c_{_{j},\sigma^{\prime}}.\label{eq:psi-Floquet-1}
\end{equation}
The hybridization now involves valence fluctuations accompanied by transitions between Floquet sectors, 
\begin{equation}
H_{hyb}=V_{0}
\sum_{m,n} \sum_{i,\sigma}f_{i\sigma}^{\dagger}\psi_{i\sigma}^{\left(m-n\right)}\left|m\right\rangle \left\langle n\right|+\text{h.c.}
\end{equation}
Effectively, the field modifies the Wannier functions, as the hybridization process is accompanied by the absorption ($m<0$) or emission ($m>0$) of photons.  This leads to symmetry-distinct form factors depending on $n$, which generate different emergent Kondo channels.  The valence fluctuation energy is also modified by the exchange of photons.
\begin{equation}
H_{atom}=\sum_{m}\left(-\left|\epsilon_{f}\right|\sum_{i,\sigma}f_{i\sigma}^{\dagger}f_{i\sigma}+m\Omega\right)\left|m\right\rangle \left\langle m\right|.
\end{equation}
The form of $g_n$ depends on the polarization, which controls the nature of the emergent Kondo channels.

\section{The effect of polarization\label{sec:polarization}}

For a generic polarization of normally incident light, the electric field, $\boldsymbol{E}\left(t\right)=\text{Re}\left[\boldsymbol{E}_{0}e^{-i\Omega t}\right]$ can be written in the circular basis, $\boldsymbol{E}_{0}=E_{+}\hat{\epsilon}_{+}+E_{-}\hat{\epsilon}_{-}$, where $E_{\pm}$ correspond to the degree of left and right circular polarization (LCP/RCP),
\begin{align}
E_{+} & =\sqrt{I}\sin\left(-\chi-\pi/4\right)e^{-i\left(\psi-\pi/2\right)},\label{eq:Eplus}\\
E_{-} & =\sqrt{I}\cos\left(-\chi-\pi/4\right)e^{i\left(\psi-\pi/2\right)}.\label{eq:Eminus}
\end{align}
\begin{figure}
\centering{}\includegraphics[width=0.8\columnwidth]{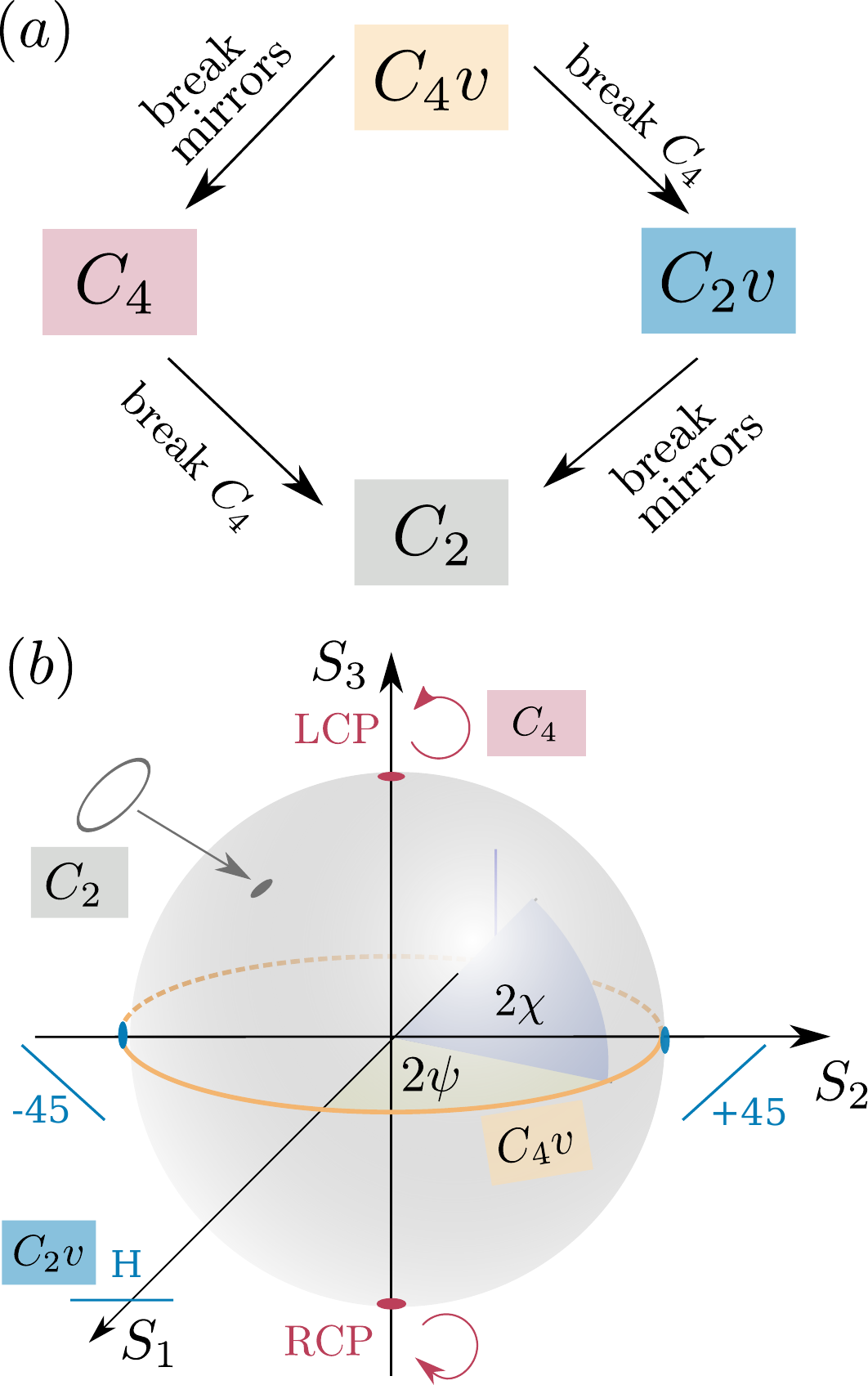}\caption{(a) Different light polarizations preserve different symmetries.  Here, we show how the original $C_{4v}$ symmetry can be broken. All the lattice symmetries are kept in the $C_{4v}$ case, leading to up to four symmetry-distinct channels. Circularly polarized light carries chirality, breaking the mirror and time-reversal symmetries and reducing the symmetry from $C_{4v}$ down to $C_{4}$. Linearly polarized light reduces the symmetry down to $C_{2}$ as the two lattice directions become nonequivalent and the mirror symmetries are generically broken except for horizontal (H), vertical (V) and $\pm 45^\circ$ directions. (b) The Poincar\'e sphere, where different points correspond to different polarizations. Different colors represent the points or paths that realize models invariant under each symmetry group shown in (a). The orange line along the equator represents an ensemble average of linearly polarized light that preserves all the original symmetries. \label{fig:Floquet-Kondo-sphere}}
\end{figure}
This parametrization allows a simple representation of the Stokes parameters~\cite{bornwolf_book,QuitoFlintPRL2021}, which indicate the degree and nature of polarization and can be written as the vector, $\boldsymbol{S}=I\left(\cos2\chi\cos2\psi,\cos2\chi\sin2\psi,\sin2\chi\right)$.
Monochromatic light is represented by $\boldsymbol{S}$ at a point on the surface of the Poincar\'e sphere, Fig.~\ref{fig:Floquet-Kondo-sphere}(b), while polarization averaging allows the Stokes vector to explore different paths on the Poincar\'e sphere.

We can now explicitly integrate the dressing of the hoppings and hybridizations, Eq.~(\ref{eq:gn_Floquet}),
\begin{align}
g_{n}\left(\boldsymbol{\delta}_{ij}\right) & =\mathcal{J}_{n}\left(A_{ij}\right)e^{in\left(\beta_{ij}+\pi\right)},
\end{align}
where $\mathcal{J}_{n}$ are Bessel functions of the first kind, and the amplitudes $A_{ij}$ are
\begin{equation}
A_{ij}=A_{0}\left|\boldsymbol{\delta}_{ij}\right|\sqrt{1+\cos2\chi\cos\left[2\left(\psi-\phi_{ij}\right)\right]}.\label{eq:A_l}
\end{equation}
The dimensionless vector potential, or fluence, $A_{0}=\frac{1}{\Omega}\sqrt{I/2}$, and $\phi_{ij}$ is the angle between the vectors $\boldsymbol{R}_{i}$ and $\boldsymbol{R}_{j}$; in this paper, we consider mostly nearest-neighbors, $|\delta_{ij}| = a = 1$.  The phase $\beta_{ij}$ is
\begin{align}
\tan\beta_{ij} & =-\text{cotan}\chi\text{cotan}\left(\psi-\phi_{ij}\right).\label{eq:tan-bl}
\end{align}
For LCP/RCP light, $\beta_{ij}=\mp\phi_{ij}$
while for linearly polarized (LP) light, $A_{ij}=\sqrt{2}A_{0}\cos\left(\psi-\phi_{ij}\right)$
and $\beta_{ij}=0$. 
The generic Wannier operator, Eq.~\eqref{eq:psi-Floquet-1} becomes
\begin{equation}
\psi_{i,\sigma}^{\left(n\right)}=\sum_{j,\sigma^{\prime}}\phi_{ij}^{\sigma\sigma^{\prime}}\mathcal{J}_{n}\left(A_{ij}\right)e^{in\left(\beta_{ij}+\pi\right)}c_{j,\sigma^{\prime}}.\label{eq:psi-Floquet}
\end{equation}
Unpolarized light is obtained by allowing the Stokes vector to trace out a path on the Poincar\'e sphere such that no symmetries are broken after averaging.  Type I light averages over the entire sphere, while type II Glauber light averages over the equator~\cite{LehnerPRA1996,Beckley2010,Shevchenko2019}; this light may be generated  most straightforwardly by combining two lasers with slightly detuned frequencies and opposite circular polarizations~\cite{Ortega-Quijano2017,QuitoFlintPRL2021}. The polarization then varies on a time scale, $T_p \gg 2\pi/\Omega = T$.  While unpolarized light is not strictly monochromatic, it allows a Floquet approach in the limit where the polarization changes sufficiently slowly~\cite{QuitoFlintPRL2021,QuitoFlintPRB2020}.  As long as the polarization time is shorter than the relaxation time of the spins ($T_{rel} \sim 2\pi/T_K$), they will feel an effective Kondo interaction given by the average over the whole path, rather than a time-dependent one.  Paths can be defined by a function $f(\psi,\chi)$, where $f(\psi,\chi) = \delta(\chi)$ for type II Glauber light, for example. The averaged Kondo coupling is then,
\begin{equation}
\left\langle J\left(\psi,\chi\right)\right\rangle =\frac{1}{\pi}\!\!\int_{0}^{\pi\!}\!\!\!d\psi\!\!\int_{-\pi/4}^{\pi/4}\!\!\!d\chi\cos2\chi f(\psi,\chi)J\left(\psi,\chi\right).
\end{equation}

\section{Emergent Kondo couplings\label{sec:Kondo_couplings}}

Our driven Anderson model consists of virtual valence fluctuations to an infinite number of excited states in different Floquet sectors.  These excited states have energies $n\Omega+U$, $n \in \mathbb{Z}$, and the valence fluctuations have form factors given by the relevant Wannier functions, $\psi_{j\sigma}^{(n)}$. Naively, each $n$ might give a different emergent Kondo channel, but inspection of the phases in Eq.~\eqref{eq:psi-Floquet} shows a periodicity with $n$ that varies based on the polarization. This periodicity is related to the reduced symmetry; $C_N$ symmetry gives periodicity $N$, and so our periodicity here is at most four.  Distinct Kondo channels are determined by the symmetry of the hybridization form factors, not the specific excited state, and different $n$ can share the same symmetry.

We therefore Fourier transform the Floquet-Wannier states,
\begin{equation}
\label{eq:wannier-ff}
    \psi_{j\sigma}^{\left(n\right)}\!= \!\sum_{\boldsymbol{k}}\!\Phi_{\boldsymbol{k};\sigma\sigma^{\prime}}^{\left(n\right)}c_{\boldsymbol{k}\sigma^{\prime}}e^{i\boldsymbol{k}\cdot\boldsymbol{R}_{j}} \!\equiv \! \sum_{\boldsymbol{k}}\!\gamma_{\Gamma_{\alpha}}^{\left(n\right)}\Phi_{\boldsymbol{k};\sigma\sigma^{\prime}}^{\Gamma_{\alpha}}c_{\boldsymbol{k}\sigma^{\prime}}e^{i\boldsymbol{k}\cdot\boldsymbol{R}_{j}},
\end{equation}
where we separate the $n$-dependent coefficient, $\gamma_{\Gamma_{\alpha}}^{\left(n\right)}$ and the $n$-independent form factor, $\Phi_{\boldsymbol{k};\sigma\sigma^{\prime}}^{\Gamma_{\alpha}}$. We identify the momentum space form-factors by the way they transform under the reduced symmetry group, labeled by the irreducible representation (irrep), $\Gamma_{\alpha}$.  The reduced symmetry group consists of all lattice symmetries preserved when the light polarization is included~\cite{altmann1994book}. For $C_{4v}$ symmetry, all possible subgroups and the associated polarization protocols are shown in Fig.~\ref{fig:Floquet-Kondo-sphere}.  LCP/RCP reduce $C_{4v}$ to $C_4$, while generic linear polarization reduces it to $C_2$.  Special linear polarizations ($\phi = 0,\pi/4,\pi/2,3\pi/4$) have $C_{2v}$ symmetry, and the average over all linear polarizations (type II Glauber) restores the $C_{4v}$ symmetry.
For the toy model, $\Phi_{\boldsymbol{k}}^{\Gamma_{\alpha}}$ is just a function of $\boldsymbol{k}$, but with spin-orbit coupling, it acquires a matrix structure in $\sigma \sigma'$. 
All irreps are allowed by symmetry, but their strength depends on the lattice structure. For instance, on the square lattice with type II Glauber light, the ``reduced'' symmetry group is still $C_{4v}$, and emergent $B_{1g}$ ($d_{x^2-y^2}$) hybridizations arise naturally from nearest-neighbor $V_0$, while $B_{2g}$ ($d_{xy}$) requires next-nearest-neighbor overlaps ($V_1$) that are typically smaller. 

A Floquet-Schrieffer-Wolff transformation is performed independently for each $n$~\cite{Schrieffer1966,Bukov2016,eckstein2017arxiv}, which generates an infinite series of Kondo interactions labeled by $n$.  We collect these into channels labeled by the irreps,
\begin{equation}
H_{FK}=\!\!\!\sum_{\substack{\boldsymbol{k},\boldsymbol{k}^{\prime},j\\\alpha,\sigma,\sigma^{\prime}}}\!\!\!J_{\Gamma_\alpha}e^{i\left(\boldsymbol{k}-\boldsymbol{k}^{\prime}\right)\cdot\boldsymbol{R}_{j}}\!c_{\boldsymbol{k}\sigma\alpha}^{\dagger}\!\left[\Phi_{\boldsymbol{k}^{\prime}}^{\Gamma_{\alpha}}\right]^{\dagger}\!\!\boldsymbol{\sigma}_{\sigma\sigma^{\prime}}\Phi_{\boldsymbol{k}}^{\Gamma_{\alpha}}c_{\boldsymbol{k}\sigma^{\prime}\alpha}\cdot\boldsymbol{S}_{j},\label{eq:HFL_toy}
\end{equation}
where $\alpha=1,\ldots,N$ labels the $N$ symmetry distinct channels and $J_0$ is the equilibrium Kondo coupling strength. $J_{\alpha}$ depends on the polarization via  $\gamma_{\Gamma_{\alpha}}^{\left(n\right)}$:
\begin{equation}
J_{\Gamma_\alpha}=J_{0}\sum_{n=-\infty}^{+\infty}\frac{\gamma_{\Gamma_{\alpha}}^{\left(n\right)}\gamma_{\Gamma_{\alpha}}^{\left(-n\right)}}{1+n\tilde{\Omega}},
\end{equation}
with $\tilde{\Omega}\equiv\Omega/\left|\epsilon_{f}\right|$. The $\gamma_{\Gamma_{\alpha}}^{\left(n\right)}$ are given in Appendix~\ref{sec:toy_model_App}, where their structure reflects the periodicity, and the amplitudes are combinations of Bessel functions of order $n$.

The full Kondo model also includes the conduction electron kinetic energy, where the hoppings are also renormalized by the Floquet field, similarly to Eq.~\eqref{eq:cond_Floquet}.  For simplicity, we assume that $\Omega \gg |t_{ij}|$, in which case, the hopping along $\boldsymbol{\delta}_{ij}$
can be approximated using a high-frequency expansion, leading to renormalized hoppings, $\tilde{t}_{ij}=\mathcal{J}_{0}\left(A_{ij}\right)t_{ij}$, where $A_{ij}$ is defined as in Eq.~\eqref{eq:A_l}~\cite{Mikami2016}. 

\section{Toy model results\label{sec:toy_model_results}}

The choice of light polarization determines the reduced symmetry group and specific emergent Kondo channels. In this section, we discuss two illustrative cases for the toy model: circular polarization, which reduces the symmetry from $C_{4v}$ to $C_4$ and breaks time-reversal without substantially splitting the ground state doublet; and the average of all linear polarizations (type II Glauber light), which preserves $C_{4v}$.  These results follow closely from Eq.~\eqref{eq:HFL_toy}, with additional details given in Appendix~\ref{sec:toy_model_App}.

\subsection{Circular polarization}

Circularly polarized light provides the simplest example, where the reduced symmetry group is $C_{4}$, as the mirror planes have been removed by the light chirality. Following the standard notation for $C_{4}$, we label the irreps (and thus the channels) $\Gamma_{i}$, $i = 1, \ldots ,4$. $\Gamma_{1}$ and $\Gamma_{2}$ transform as $s$-wave and $d$-wave, respectively, where $d_{xy}$ and $d_{x^2-y^2}$ are indistinguishable in $C_4$.  $\Gamma_{3}$ and $\Gamma_{4}$ are chiral, and transform like $p_x\pm ip_y$. 
The full Floquet-Wannier functions, $\psi^{(n)}_i$ are given in the Appendix~\ref{sec:toy_model_App}, where $A_{ij}=A_{0}\left|\boldsymbol{\delta}_{ij}\right|$ and the phases only differ for $n \mod 4$. $n=0$ gives $\Gamma_1$, $n=2$ gives $\Gamma_2$ and $n = 1,3$ give $\Gamma_{3,4}$, as shown in Fig. \ref{fig:Floquet_Kondo_schem}(c).  We therefore find four distinct Kondo couplings, one for each $\Gamma_i$. For nearest-neighbor hybridization, these depend on the frequency, $\tilde{\Omega}=\Omega/|\epsilon_f|$ and fluence, $A_0$, as follows,
\begin{align}
J_{\Gamma_{1}} & =J_{0}\sum_{m=-\infty}^{\infty}\frac{\mathcal{J}_{4m}^{2}\left(A_{0}\right)}{1+4m\tilde{\Omega}},\nonumber \\
J_{\Gamma_{3}} & =J_{0}\sum_{m=-\infty}^{\infty}\frac{\mathcal{J}_{4m+1}^{2}\left(A_{0}\right)}{1+\left(4m+1\right)\tilde{\Omega}},\nonumber \\
J_{\Gamma_{2}} & =J_{0}\sum_{m=-\infty}^{\infty}\frac{\mathcal{J}_{4m+2}^{2}\left(A_{0}\right)}{1+\left(4m+2\right)\tilde{\Omega}},\nonumber \\
J_{\Gamma_{4}} & =J_{0}\sum_{m=-\infty}^{\infty}\frac{\mathcal{J}_{4m+3}^{2}\left(A_{0}\right)}{1+\left(4m+3\right)\tilde{\Omega}}.\label{eq:toy-circular-couplings}
\end{align}
The form factors are $\Phi^{\Gamma_i}_{\bf k} = \{\cos k_x + \cos k_y,  \cos k_x - \cos k_y, \sin k_x + i \sin k_y, \sin k_x - i \sin k_y \}$ for $i = 1,2,3,4$, and the four channel Kondo model is given by Eq. (\ref{eq:HFL_toy}).

\begin{figure}
\includegraphics[width=1\columnwidth]{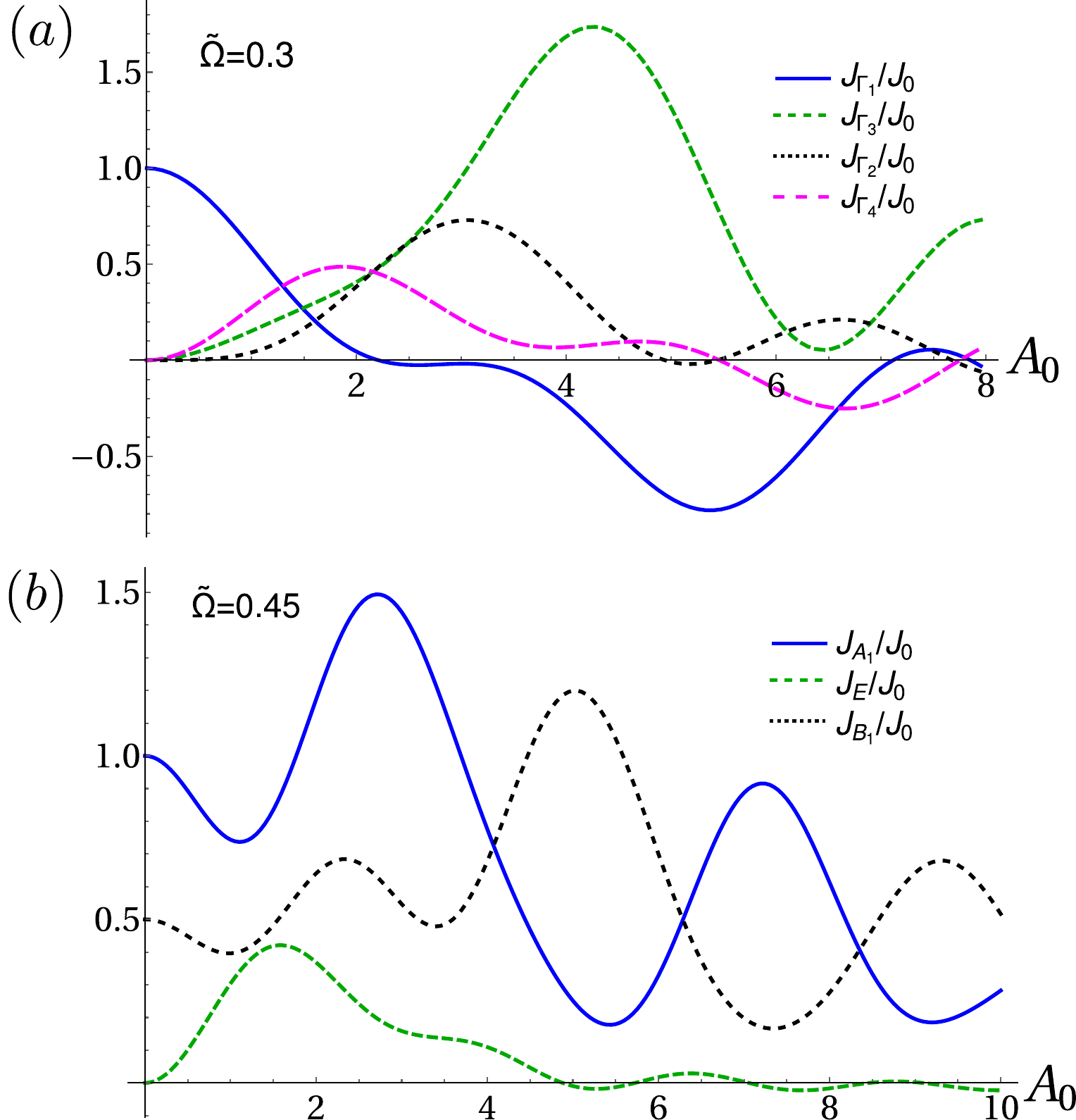}

\caption{Variation of the Kondo couplings, given in units of $J_{0}=4V_{0}^{2}/\left|\epsilon_{f}\right|$ as functions of the dimensionless vector potential strength (fluence), $A_{0}$. (a) Left circularly polarized light leads to four channels, $s\,\left(\Gamma_{1}\right),p_{+}\,\left(\Gamma_{3}\right),d\,\left(\Gamma_{2}\right)$ and $p_{-}\,\left(\Gamma_{4}\right)$. For $\tilde{\Omega}=\Omega/\left|\epsilon_{f}\right|=0.3$ and $A_{0}\approx2.15$, the three new channels are exactly degenerate, while the coupling of the initial $s$-wave channel is very close to zero. Changing from left to right circularly polarized light simply exchanges $J_{\Gamma_{3}}\rightleftarrows J_{\Gamma_{4}}$. (b) By averaging over linear polarization, we restore the full square-lattice symmetry, leading to four channels. Here, we assume that the initial Kondo coupling of the $A_1$(s wave) channel was twice the $B_1$ ($d_{x^2-y^2}$). The $E$ channel is exactly degenerate, as $p_{x}$ and $p_{y}$ have the same Kondo coupling, as required by symmetry. For this case, we choose $\tilde{\Omega}=0.45$. \label{fig:Coupling-values}}

\end{figure}

These Kondo couplings become large near the resonances $\tilde{\Omega} \approx -1/n$, which means different $\tilde{\Omega}$ can enhance different channels, although very close to the resonance heating becomes problematic.  In Fig.~\ref{fig:Coupling-values}(a), we plot the strength of each channel for $\tilde{\Omega}=\Omega/\left|\epsilon_{f}\right|=0.3$.
For small $A_{0}$, the dominant channel is $\Gamma_{1}$, the extended s-wave, little changed from the static case. As $A_{0}$ increases, $J_{\Gamma_{1}}$ is reduced while the other channels are initially enhanced although all channels are eventually suppressed for sufficiently large $A_0$ due to the Bessel functions in their amplitudes. Here, we have chosen the specific $\tilde{\Omega}$ such that there is an exact degeneracy of three different channels for a specific $A_0 = 2.15$. 
Four degenerate channels can be found with similar fine-tuning when the polarization average restores the $C_{4v}$ symmetry (see Appendix~\ref{sec:toy_model_App}). The Kondo couplings may become negative, depending upon $A_0$ and $\tilde{\Omega}$, as seen for $J_{\Gamma_{1}}$ here; negative Kondo couplings are irrelevant, making it possible to filter out certain channels entirely.
Tuning $\tilde{\Omega}$ enhances the different channels, with further examples given in Appendix~\ref{sec:toy_model_App}.

The $C_{4v}$ symmetry may be restored by averaging over LCP and RCP polarizations, which is straightforward to do theoretically by simply adding the two polarization cases and identifying the channels. The two chiral irreps, $\Gamma_3$ and $\Gamma_4$ mix to give the doubly degenerate $E$ representation ($p_x, p_y$), which now has $J_E = 1/2(J_{\Gamma_3}+J_{\Gamma_4})$ with contributions from all odd $n$. This term automatically gives a perfectly degenerate two-channel Kondo effect if it is dominant. $J_{\Gamma_1}$ is identified with the $A_1$ channel and $J_{\Gamma_2}$ with the $B_1$ ($d_{x^2-y^2}$) channel.  There are two additional possible channels for $C_{4v}$, which involve further neighbor hybridizations: $B_{2}$ ($d_{xy}$) is only present with next-nearest-neighbor hybridizations, $V_1$, and is suppressed by a factor of $V_1^2/V_0^2$ as compared to nearest neighbors; and $A_2$ ($g_{xy (x^2-y^2)}$) requires even further neighbor hybridizations and is unlikely to be realized. Note that $A_2$ can also have a $p_z$ component, which cannot be altered by normally incident light and is not relevant for our strictly two-dimensional model.

\subsection{Average of linear polarizations}

We next consider slowly varying linear polarization, corresponding
to averaging over the equator of the Poincar\'e sphere.  The net effect of this averaging is again to restore the full lattice symmetry to $C_{4v}$; as long as the polarization changes sufficiently slowly ($T_p \gg T$), time-reversal symmetry is also preserved. 
The Wannier functions for linear polarization, the resulting Kondo couplings, and the effect of averaging are given in Appendix~\ref{sec:toy_model_App}, as these are fairly complicated.  The final channels are identical to those found for the LCP/RCP average, due to the $C_{4v}$ symmetry, but here, odd $n$ leads to degenerate $E$ ($p$-wave) channels, while all even $n$ contribute to both $A_1$ ($s$), $B_1$($d_{x^2-y^2}$) and $B_2$ ($d_{xy}$) channels, where the $B_2$ channel is again only present with further neighbor hybridization $V_1$ (see Appendix~\ref{sec:toy_model_App}).  

Fig.~\ref{fig:Coupling-values}(b) shows how these channel strengths evolve with fluence, $A_0$.  Motivated by the possibility of enhancing composite pairing, we take two non-degenerate equilibrium Kondo channels: $s$-wave ($A_1$) and $d_{x^2-y^2}$ ($B_1$), with initial values of $J_{A_1} = J_0$, $J_{B_1} = J_0/2$, or alternately, $|\epsilon_{f,A_1}| = |\epsilon_{f}|$ and $|\epsilon_{f,B_1}| = 2 |\epsilon_f|$.    Here, the initial $d$-wave channel is also modified by the Floquet phases to contribute to the $s$ and $p$ wave channels with finite fluence, leading to the enhancement of the $A_1$ channel with intermediate fluences.  For this choice of $\tilde{\Omega}$, $|\epsilon_{f,A_1}|$, and $|\epsilon_{f,B_1}|$, the doubly degenerate ($E$) interaction is always the smallest, and can be neglected~\cite{KimCox_PRB_1997}, but for other initial conditions it may be the largest Kondo coupling for some range of fluence, and there will be three channel Kondo degeneracies with the $A_1$ or $B_1$ channels, and four channel degeneracies at some extremely fine tuned point.  Note that $B_2$ ($d_{xy}$) and $A_2$ symmetries are not included, as we restrict our initial hybridization to nearest-neighbors.  These toy model results capture much of the relevant physics for more realistic cases, as we consider in the next two sections.

\section{Quantum dot realizations\label{sec:quantum_dots}}

Quantum dots have long been used to study multi-channel Kondo impurity physics~\cite{Goldhaber-Gordon_Nature_1998,Potok_Nature_2007,Pierre_Science_2018}, and in this section we discuss how our impurity toy model results can be mapped onto a quantum dot model for both arbitrary fixed and averaged polarizations. We consider a two-dimensional generalization of Ref.~\cite{eckstein2017arxiv}.  There, the quantum dot was coupled to two leads (L,R), and subject to an oscillating asymmetric bias, $\Delta_{0}\sin\Omega t\sum_{i}\left(c_{Li}^{\dagger}c_{Li}-c_{Ri}^{\dagger}c_{Ri}\right)$, which led to an emergent $p$-wave hybridization.  Our two-dimensional analog has four leads, with similar oscillating biases across both left (L) to right (R) and top (T) to bottom (B) leads, leading to several distinct channels depending on the driving protocol, in complete analogy to the toy model impurity. The time-dependent Hamiltonian is 

\begin{align}
H & \left(t\right)=-t_{1}\sum_{i\eta\sigma}\left[c_{i+1\eta\sigma}^{\dagger}c_{i\eta\sigma}+\text{h.c.}\right]+H_{hyb}\left(0\right)+\nonumber \\
 & +H_{dr}\left(t\right)-\left|\epsilon_{f}\right|f_{0\sigma}^{\dagger}f_{0\sigma}+U\hat{n}_{f0\uparrow}\hat{n}_{f0\downarrow},\label{eq:dot_time_dependent}
\end{align}
where $\psi_{0\sigma} = \sum_\eta c^\dagger_{0\eta\sigma}$, $\eta =L,R,T,B$,

\begin{equation}
H_{hyb}\left(0\right)=V_{0}\sum_{\sigma}\left[f_{0\sigma}^{\dagger}\psi_{0\sigma}+\text{h.c.}\right]
\end{equation}
is a time-independent hybridization and $t_{1}$ is the c-electron hopping. We consider a driving chemical potential that is different for each lead, and parametrized as $\mu_{\eta}\left(t\right)=\boldsymbol{E}\left(t\right)\cdot\boldsymbol{\delta}_{\eta}$. We define $\boldsymbol{\delta}_{\eta}$ as the unit vector along the wire labeled by $\eta$, leading to

\begin{equation}
H_{dr}\left(t\right)=\sum_{\eta=\left\{ R,L,T,B\right\} }\left[\boldsymbol{E}\left(t\right)\cdot\boldsymbol{\delta}_{\eta}\right]c_{\eta\sigma}^{\dagger}c_{\eta\sigma}.
\end{equation}
The notation $\boldsymbol{E}\left(t\right)$ anticipates the direct analogy with the toy model. The quantum dot may be mapped into this impurity problem using a time-dependent unitary transformation,

\begin{equation}
\tilde{H}\left(t\right)=e^{-iS\left(t\right)}\left[H\left(t\right)-i\partial_{t}\right]e^{iS\left(t\right)}.\label{eq:H_rotation}
\end{equation}
Here, $\tilde{H}\left(t\right)$ is the transformed time-dependent Hamiltonian
and $S\left(t\right)$ is the generator of the transformation. We can eliminate $H_{dr}\left(t\right)$ from $\tilde{H}\left(t\right)$ using $S\left(t\right)=-\int^{t}dt^{\prime}H_{dr}\left(t^{\prime}\right)$, as can be seen from the time derivative in Eq.~(\ref{eq:H_rotation}). Only the hybridization and conduction electron hopping terms are affected by this unitary transformation. The $c$-electrons transform as  $e^{-iS\left(t\right)}c_{i\eta\sigma}e^{iS\left(t\right)}=e^{i\boldsymbol{A}\left(t\right)\cdot\boldsymbol{\delta}}c_{i\eta\sigma}$, leading to a time-dependent hybridization,

\begin{equation}
H_{hyb}\left(t\right)=V_0\sum_{\eta=\left\{ L,R,T,B\right\} }e^{-i\boldsymbol{A}\left(t\right)\cdot\boldsymbol{\delta}_{\eta}}c_{0\eta\sigma}^{\dagger}f_{0\eta\sigma}.
\end{equation}
The kinetic energy term is also modified by a phase, $c_{i+1\eta\sigma}^{\dagger}c_{i\eta\sigma}\rightarrow c_{i+1\eta\sigma}^{\dagger}e^{-i\boldsymbol{A}\left(t\right)\cdot\boldsymbol{\delta}_{\eta}}c_{i\eta\sigma}$. As before, we assume that the oscillation frequency, $\Omega$ is large compared with the conduction band bandwidth, $D$, leading to the same simple renormalization of the conduction electron bands.

The results found above for circular polarization can be reproduced with $\boldsymbol{E}_{\pm}\left(t\right)=E_{0}\left(\cos\Omega t,\pm\sin\Omega t\right)$, where $\pm$ correspond to left and right circular polarizations. The four leads share an oscillation frequency, $\Omega$, but with relative phases between each lead of $\pm \pi/4$. This set-up leads to the dynamically-generated four-channel Kondo model discussed above. 

Polarization averaging can be reproduced by applying two detuned oscillating biases in analogy to creating type II Glauber light by applying two detuned lasers~\cite{QuitoFlintPRL2021},  taking

\begin{align}
    \boldsymbol{E}(t) & = E_0 \left(\begin{array}{c}\cos \Omega_p t\\\sin \Omega_p t\end{array}\right)\cos{\Omega t},\cr 
    & = \frac{E_0}{2} \text{Re}\left[ \left(\begin{array}{c}1\\i\end{array}\right)\mathrm{e}^{-i \Omega_+ t}+ \left(\begin{array}{c}1\\-i\end{array}\right)\mathrm{e}^{-i \Omega_- t} \right]. \label{eq:quasi-mon-protocol}
\end{align}
with $\Omega_{\pm}=\Omega\pm\Omega_{p}$. For $N=\Omega/\Omega_{p} \gg 1$, the four-channel Kondo model is realized, identically to the average over linear polarizations discussed above.

Finally, we discuss a third example. By choosing $\boldsymbol{E}\left(t\right)$ to be linearly polarized along the diagonal ($\psi = \pi/4, 3\pi/4$), we find an effective three channel Kondo model. The chemical potential is $\mu=\pm \mu_{0} \cos\Omega t$, where $+(-)$ corresponds to the L,T (R,B) wires. If we had additional leads analogous to further neighbors (e.g. - $V_1 \neq 0$), the reduced symmetry group would be $C_{2v}$, but here the structure restores the $C_{4v}$ symmetry. There is an $s$-wave ($A_1$) channel for $n$ even and degenerate $p_x$ and $p_y$ ($E$) channels for $n$ odd: $\Phi_{\boldsymbol{k}}^{\left(n\,\text{e}\right)}=2\mathcal{J}_{n}\left(A_{0}\right)\left(\cos k_{x}+\cos k_{y}\right)$ and $\Phi_{\boldsymbol{k}}^{\left(n\,\text{o}\right)}=2i\mathcal{J}_{n}\left(A_{0}\right)\left(\sin k_{x}+\sin k_{y}\right)$; see Appendix~\ref{sec:toy_model_App} for details.  

\section{Realistic C\lowercase{e} model \label{sec:Ce_model}}

Our simple toy model assumed Heisenberg local moments that were decoupled from the lattice and had an initial $s$-wave hybridization. In real materials, the $f$-electrons forming the local moments are strongly spin-orbit coupled ($J=5/2$ for Ce). Their ground states are described by \emph{double-group} irreps, where the hybridization symmetry is determined by the Wannier functions that add or remove an $f$-electron from these ground state doublets; these symmetries are captured by the double group irreps of the reduced symmetry group. Motivated by the Ce 115 materials~\cite{Petrovic_2001,Park_Nature_2006,Flint_NatPhys_2008,Singh_2015,Prozorov_PRL_2015}, we consider a two-dimensional square lattice where the $J=5/2$ Ce ions have a $\Gamma_7^+$ ground state, $|\Gamma_7^+\pm\rangle = \cos \xi |\pm 5/2\rangle + \sin \xi |\mp 3/2\rangle$, written in terms of the $J_z$ configurations, where $\xi$ is a materials dependent angle. These Ce hybridize with $s$-wave conduction electrons sitting on the same lattice, where the Wannier states are still constructed by a superposition of conduction electrons on neighboring sites, but now with a spin-dependent form-factor~\cite{AlexandrovColeman_PRL_2013}.
In the following, we define the doublet of Wannier functions, $\psi_{\boldsymbol{R}_{j}}^{T}\equiv\left(\psi_{\Gamma_{7}^+,\boldsymbol{R}_{j},+},\psi_{\Gamma_{7}^+,\boldsymbol{R}_{j},-}\right)$.
In momentum space for Floquet sector $n$, the hybridization matrix is purely off-diagonal and $\Phi_{\boldsymbol{k}}^{\left(n\right)}$ is 

\begin{equation}
\Phi_{\boldsymbol{k}}^{\left(n\right)}=\left(\begin{array}{cc}
0 & f^{\left(n\right)}\left(\boldsymbol{k}\right)\\
\left(-1\right)^{n+1}\left[f^{\left(-n\right)}\left(\boldsymbol{k}\right)\right]^{*} & 0
\end{array}\right).
\end{equation}
The functions $f^{\left(n\right)}\left(\boldsymbol{k}\right)$ depend
on the polarization, and are shown in Appendix \ref{sec:Ce_model_App}.  Identically to the toy model, the symmetry is reduced from $C_{4v}$ to $C_{4}$ for LCP or RCP, and to $C_{2}$ for LP. 
Here, we simply quote the results for the ensemble of LP and leave the details and other polarization choices to Appendix~\ref{sec:Ce_model_App}.
\begin{figure}
\begin{centering}
\includegraphics[width=1\columnwidth]{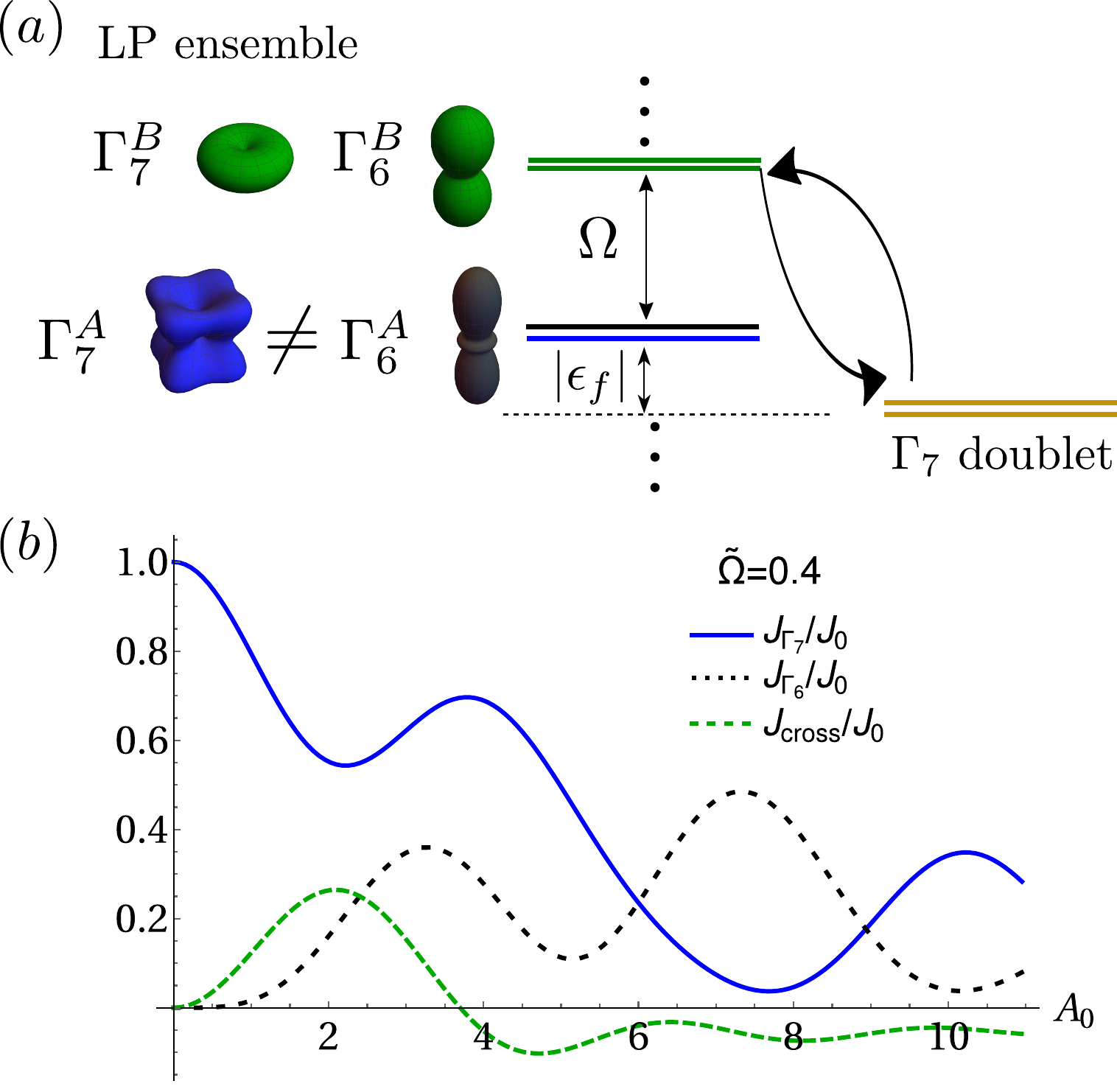}
\par\end{centering}
\caption{Ce model level structure and hybridization profiles for light averaged over an ensemble of linear polarization. The $A,B$ labels correspond to two different hybridizations of the same irrep, formed from $J=5/2$ and $J=3/2$ states of Ce, respectively.
\label{fig:Level-structure-Ce}}
\end{figure}

To understand the relevant symmetries, we look to the modified valence fluctuations of Ce coupled to an ensemble of linearly polarized light, as shown in Fig.~\ref{fig:Level-structure-Ce}(a).  Once the $C_{4v}$ symmetry is restored, there are four channels, 
\begin{align}
\Phi_{7,\boldsymbol{k}}^{A} & =\left(\begin{array}{cc}
0 & i\sin k_{x}-\sin k_{y}\\
i\sin k_{x}+\sin k_{y} & 0
\end{array}\right),\nonumber \\
\Phi_{7,\boldsymbol{k}}^{B} & =\left(\begin{array}{cc}
\cos k_{x}-\cos k_{y} & 0 \\
0 & -(\cos k_{x}-\cos k_{y})
\end{array}\right),\nonumber \\
\Phi_{6\boldsymbol{k}}^{A} & =\left(\begin{array}{cc}
0 & i\sin k_{x}+\sin k_{y}\\
i\sin k_{x}-\sin k_{y} & 0
\end{array}\right),\nonumber \\
\Phi_{6\boldsymbol{k}}^{B} & =\left(\begin{array}{cc}
\cos k_{x}+\cos k_{y} & 0 \\
0 & -(\cos k_{x}+\cos k_{y})
\end{array}\right).\label{eq:C4v-matrices}
\end{align}
Two of the channels are familiar from equilibrium: 7A is the symmetry of this equilibrium model, while the new 6A channel involves a $J=5/2$ conduction electron Wannier function in $\Gamma_6$, $|\Gamma_6\pm\rangle =|\pm 1/2\rangle$.  Odd $n$ fluctuations lead to two new channels: 7B and 6B, which are $J=3/2$ versions of these irreps. The change in total angular momentum is due to the addition of an odd number of $L=1$ photons. The two A channels are both found for even $n$ and give the Kondo interactions, $J_{7A}$ and $J_{6A}$,
\begin{align}
J_{7A} & =\frac{J_{0}}{2}\sum_{m=-\infty}^{\infty}\frac{A_{2m}+B_{2m}}{1+2m\tilde{\Omega}},\nonumber \\
J_{6A} & =\frac{J_{0}}{2}\sum_{m=-\infty}^{\infty}\frac{A_{2m}-B_{2m}}{1+2m\tilde{\Omega}},
\end{align}
where the $A_n$, $B_n$ coefficients are averages over polarization-dependent Bessel functions given in  Appendix~\ref{sec:Ce_model_App}.
By contrast, the B channels give a single unusual, anisotropic, Kondo-like interaction that we call $J_{cr}$ due to the presence of cross-channel terms,
\begin{equation}
J_{\text{cr}}=\frac{J_{0}}{2}\sum_{m=-\infty}^{\infty}\frac{A_{2m+1}}{1+\left(2m+1\right)\tilde{\Omega}}.
\end{equation}
The full form of the interaction is given by,
\begin{align}
H_{FK}& = \!\!\!\sum_{\boldsymbol{k},\boldsymbol{k}^{\prime},j}\!\!\left\{\!\left[J_{7A}\left(\Phi_{7,\boldsymbol{k}}^{A}\right)^{\dagger}\!\!\boldsymbol{\sigma}\Phi_{7,\boldsymbol{k}^{\prime}}^{A} +J_{6A}\left(\Phi_{6,\boldsymbol{k}}^{A}\right)^{\dagger}\!\!\boldsymbol{\sigma}\Phi_{6,\boldsymbol{k}^{\prime}}^{A}\right]\!\cdot\! \boldsymbol{S}_j\right.\cr
&\! \!\!\!\!\!\!+J_{cr}\!\left[\left(\Phi_{6,\boldsymbol{k}}^{B}\right)^{\dagger}\boldsymbol{\sigma}_\perp\Phi_{7,\boldsymbol{k}^{\prime}}^{B}+\left(\Phi_{7,\boldsymbol{k}}^{B}\right)^{\dagger}\boldsymbol{\sigma}_\perp\Phi_{6,\boldsymbol{k}^{\prime}}^{B}\right]\cdot \boldsymbol{S}_{j\perp}\cr
&\! \!\!\!\!\!\!\left.+J_{cr}\! \left[\left(\Phi_{6,\boldsymbol{k}}^{B}\right)^{\dagger}\!\!\sigma_z\Phi_{6,\boldsymbol{k}^{\prime}}^{B}+\left(\Phi_{7,\boldsymbol{k}}^{B}\right)^{\dagger}\!\!\sigma_z\Phi_{7,\boldsymbol{k}^{\prime}}^{B}\right]\!S_{jz}\right\}\!e^{i(\boldsymbol{k}-\boldsymbol{k}^{\prime}\!)\cdot\boldsymbol{R}_{j}}
\label{eq:HFL_Ce}
\end{align}
where we have suppressed the $\sigma,\sigma'$ indices.  The first line gives the expected Kondo effects in the 7A and 6A channels, while the second line gives an antiferromagnetic interaction between the perpendicular components of $\boldsymbol{S}_{j}$ and an inter-channel conduction electron spin density involving both 6B and 7B electrons.  Finally the third term gives an Ising Kondo interaction where the $S_{j,z}$ local moment component is screened by the $z$ component of both 6B and 7B conduction electron spin densities, with guaranteed channel degeneracy.  The $J_{cr}$ interaction has not been previously studied, although the anisotropic Kondo impurity has been studied and flows to the isotropic point, suggesting naturally degenerate two-channel Kondo behavior when $J_{cr}$ is dominant.  This term is generically present, but if desired, the appropriate choice of frequency, fluence and polarization protocol can ensure that it is the smallest of the three couplings and thus irrelevant, as seen in  Fig.~\ref{fig:Level-structure-Ce}(b).

\section{Composite order\label{sec:Composite_order}}

While multi-channel Kondo impurities lead to the fascinating possibility of non-Abelian anyons~\cite{Lopes_PRB_2020,Komijani_PRB_2020,EmeryKivelson_PRB_1992}, the multi-channel Kondo lattice is an open problem. The possible phases of a multi-channel Kondo lattice~\cite{CoxZawadowski1998} range from magnetic and multipolar orders, where the conduction electrons decouple from the local moments, to non-Fermi liquids~\cite{cox96}, symmetry-breaking heavy Fermi liquids (hastatic orders)~\cite{schauerte05,hoshino2011,ChandraFlint2013}, and finally, composite pair superconductivity~\cite{Coleman_PRB_1999,Flint_NatPhys_2008,hoshino2014}. Composite pair superconductivity is particularly interesting, as it does not require exact channel degeneracy, and it has been proposed to be relevant for the 115 family of heavy fermion superconductors~\cite{Flint_NatPhys_2008,flint10}.  In composite pairing, the heavy Cooper pairs are formed by combining a spin triplet of two conduction electrons in orthogonal channels with a local moment spin flip to form an overall singlet pair.  In the 115 materials, it has been proposed that a combination of $\Gamma_7^+$ and $\Gamma_6$ channels leads to d-wave composite pairs that coexist with and reinforce the d-wave pairing mediated by magnetic fluctuations. As the $\Gamma_6$ channel involves virtual valence fluctuations to the higher energy 4f$^2$ excited state, the two channels are not close to degenerate and the degree of composite pairing is expected to be relatively weak. However, the maximum superconducting transition temperature, $T_c$ can be as large as the Kondo temperature for equal channel strengths, which implies that Floquet engineering could be used to enhance the superconductivity in these materials.

In this section, we focus on the possibility and nature of composite pairing in our emergent multi-channel Kondo Hamiltonians, and neglect other ordered phases.  The two strongest channels determine the nature of the composite pairs, which are always spin-singlet for antiferromagnetic Kondo interactions, but can be either even-parity and even-frequency or odd-parity and odd frequency depending on the channel combination.  To understand the nature of the superconducting order parameter, we turn to the mean-field treatment of composite pairing in the two-channel Kondo lattice.

We first rewrite the Floquet-Kondo lattice interaction in real space,
\begin{equation}
H_{FK} = \sum_{j\alpha,\sigma,\sigma^{\prime}}\!J_{\Gamma_\alpha}\left(\psi^\dagger_{j\sigma\alpha}\boldsymbol{\sigma}_{\sigma\sigma^{\prime}}\psi_{j\sigma^{\prime}\alpha}\right)\cdot\boldsymbol{S}_{j},
\end{equation}
where $\psi_{j\sigma \alpha}= \sum_{\boldsymbol{k}\tau}\Phi_{\boldsymbol{k};\sigma\tau}^{\Gamma_{\alpha}}c_{\boldsymbol{k}\tau}e^{i\boldsymbol{k}\cdot\boldsymbol{R}_{j}}$.  We now introduce the fermionic symplectic-$N$ spin representation, $S_{\alpha \beta}(j) = f^\dagger_{j\alpha} f_{j\beta} - \mathrm{sgn}(\alpha \beta)f_{j-\beta}^\dagger f_{j-\alpha}$, where we now have $N$ types of spins, $\alpha \in \{-N/2,\ldots,N/2\}$, for both  the $c$- and $f$-electrons, generalizing $SU(2) \cong SP(2)$ to $SP(N)$. This symplectic-$N$ representation retains the time-reversal properties of $SU(2)$ for all even $N$, enabling a controlled mean-field treatment of superconductivity~\cite{Flint_NatPhys_2008}.  The resulting quartic Hamiltonian,
\begin{align}
H_{FK} = -\!\!\sum_{j,\alpha}\!\frac{J_{\Gamma_\alpha}}{N}\!\left[(\psi^\dagger _{j\alpha}f_{j}) 
(f^\dagger_j\psi_{j\alpha})+(\psi ^\dagger_{j\alpha} \epsilon^\dagger f^*_j)(f_{j}^T 
\epsilon \psi _{j\alpha})
\right]
\end{align}
has two terms per channel.  For conciseness, we have suppressed the spin indices by writing $\psi$ and $f$ as vectors of length $N$; $\epsilon$ is an antisymmetric large-$N$ generalization of $i \sigma_2$, and $f^*_j = \left(f^\dagger\right)^T$.   In order for this spin representation to reproduce only the physical Hilbert space, it must be accompanied by a constraint that fixes the $f$-occupation on each site to half-filling. This constraint is enforced by a vector of Lagrange multipliers, $\vec{\lambda}_j$.
All quartic terms can be decoupled by Hubbard-Stratonovich fields, leading to normal, $V_{\Gamma_\alpha} \propto \langle \psi^\dagger_{\alpha} f\rangle$ and anomalous, $\Delta_{\Gamma_\alpha} \propto \langle \psi^\dagger_{\alpha}\epsilon f^\dagger \rangle$ hybridizations in each Kondo channel. We will focus on the two-channel case, and call these two channels $\Gamma_1$ and $\Gamma_2$.  These two labels may refer to either two different representations (e.g. - $A_1$ and $B_1$) or two components of the same irrep ($p_x$ and $p_y$ for $E$), in which case channel degeneracy, $J_{\Gamma_1} = J_{\Gamma_2}$ is guaranteed. 

This Hamiltonian possesses an $SU(2)$ gauge symmetry, $f \longrightarrow uf+ v\epsilon^\dagger f^\dagger$, which may be used to eliminate the anomalous term in the first channel ($\Delta_1$). Composite pair superconductivity occurs when the product $V_1 \Delta_2 \sim \langle c^\dagger_{1} c^\dagger_{2}\epsilon f^\dagger f\rangle$ is nonzero~\cite{Flint_NatPhys_2008}.   The mean-field values of all fields, including constraint fields, may be calculated using the saddle point approximation, which is exact as $N\rightarrow \infty$.  We assume that all mean-field parameters are spatially uniform, and find that $V_2$ is typically zero, while both $V_1$ and $\Delta_2$ will develop at different temperatures.  In order to concisely write the mean-field Hamiltonian, we introduce Nambu notation,
$\tilde{c}^\dagger _{\bk }= (c^\dagger_{\bk},\epsilon c_{-\bk})$, 
$\tilde{f}^\dagger _{\bk }= (f^\dagger_{\bk},\epsilon f_{-\bk})$,
and define $\mathcal{V}_k = V_{1} \Phi^{\Gamma_1}_{\bk} \tau_3 +\Delta_{2} \Phi^{\Gamma_2}_{\bk} \tau_1$.  The mean-field Hamiltonian is now,
\begin{align}
\label{fullH}
H &= \sum_k 
\left(
\begin{array}{cc}
\tilde{c}_{k}^\dagger & \tilde{f}_{k}^\dagger
\end{array}
\right)
\left[
\begin{array}{cc}
\epsilon_k \tau_3 & \mathcal{V}_k^\dagger \\
\mathcal{V}_k & \lambda \tau_3 
\end{array}
\right]
\left(
\begin{array}{c}
\tilde{c}_{k} \\
\tilde{f}_{k}
\end{array}
\right)\nonumber\\
&+ N\left(\frac{V_{1}^\dagger V_{1}}{J_1}+\frac{\Delta_{2}^\dagger \Delta_{2}}{J_2}\right),
\end{align}
where $\lambda$ is the remaining non-zero Lagrange multiplier enforcing the constraint $n_f = N/2$.  The resulting free energy is minimized to find the values of $V_1$, $\Delta_2$ and $\lambda$ as a function of temperature, as well as the Kondo temperature, $T_K$ where $|V_1|+|\Delta_2|$ turns on and superconducting transition temperature, $T_c$, where $|V_1 \Delta_2|$ turns on.  The superconducting gap function is proportional to 
\begin{equation}
\Delta\left(\boldsymbol{k}\right)\propto\text{Tr}\left[\Phi_{2\boldsymbol{k}}^{\dagger}\Phi_{1\boldsymbol{k}}\right].\label{eq:gap_SC}
\end{equation}
In the toy model, the gap function is the product of the two hybridizations of the largest channels. For $\Gamma_1 = A_1$ and $\Gamma_2 = B_1$, the resulting superconductivity is $d_{x^2-y^2}$. The $E$ case is special, as both  $\Delta_{xy}\left(\boldsymbol{k}\right)\propto\sin k_{x}\sin k_{y}$ and $\Delta_{x^{2}-y^{2}}\left(\boldsymbol{k}\right)\propto\cos k_{x}-\cos k_{y}$ are possible for different basis choices ($p_x$, $p_y$ versus $p_x\pm i p_y$), meaning the superconducting order parameter has multiple components and will be highly sensitive to the bandstructure.  The combination of $A_1$ or $B_1$ and $E$ can lead to three degenerate channels: the possible composite order parameters include those from just $E$ as well as odd-frequency $p$ or $f$-wave composite order coming from combining the $p_{x,y}$ $E$ hybridizations with the $s$ or $d$-wave $A_1$ or $B_1$ hybridizations. In general, the possible phase space is very rich, with transitions between different superconducting symmetries as the fluence or frequency are tuned.

The Ce model is more difficult to interpret due to the unusual form of $J_{cr}$, although if $J_{7A}$ and $J_{6A}$ dominate, the resulting superconductivity will be $d_{x^2-y^2}$, as found in previous composite pairing studies~\cite{flint10}.  To get an idea of the possibilities, we can consider $SU(2)$ symmetric Kondo interactions in all four possible channels (6A,7A,6B,7B).  6A x 7A and 6B x 7B both give $d_{x^2-y^2}$ pairing, while any combination that mixes A and B irreps gives odd-frequency $p$ or $f$-wave pairing, due to the additional unit of angular momentum.

\begin{figure}
\begin{centering}
\includegraphics[width=1\columnwidth]{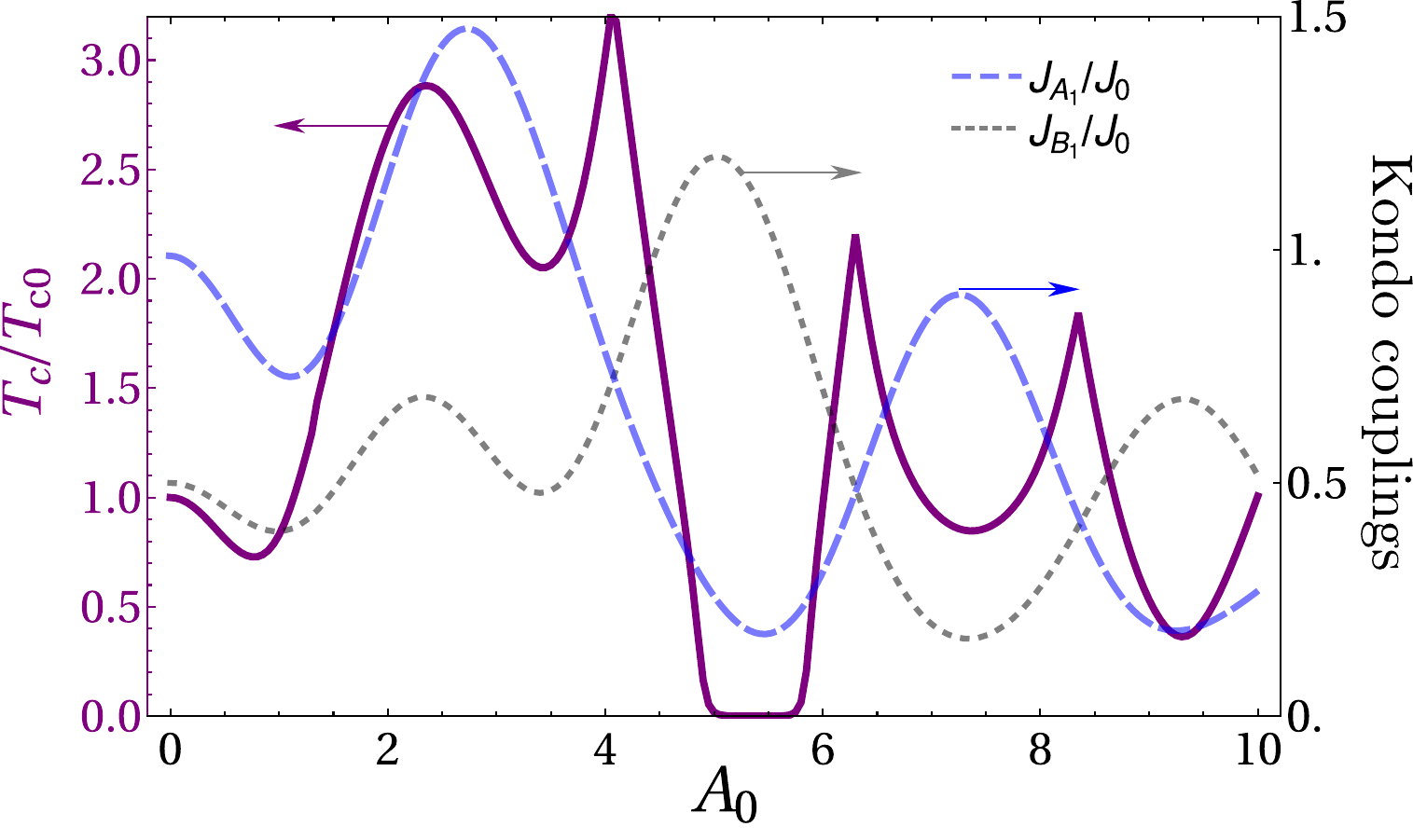}
\par\end{centering}
\caption{Superconducting transition temperature versus fluence for the toy model, with parameters given by Fig.~\ref{fig:Coupling-values}(b) [$\tilde{\Omega}=0.3$]. $T_c$ is normalized by the equilibrium value, $T_{c0}$, and shows up to a three-fold enhancement.  \label{fig:toy_model_Tc}}
\end{figure}

We explicitly treat the nearest-neighbor square lattice toy model where the polarization is averaged over all linear polarizations, for the case shown in Fig. \ref{fig:Coupling-values}(b).  Here, the $J_{A_1}$ and $J_{B_1}$ channels always dominate the doubly degenerate $E$ channel, which we therefore neglect as irrelevant.  We set $\Phi_{1\boldsymbol{k}} = \cos k_x + \cos k_y$ and $\Phi_{2\boldsymbol{k}} = \cos k_x - \cos k_y$, and the resulting superconductivity is d-wave in nature ($d_{x^2-y^2}$). We plot the superconducting transition temperature as a function of fluence in Fig. \ref{fig:toy_model_Tc}, where the conduction electron hopping $t_{1}(A_0)$ and Kondo couplings $J_{A_1}(A_0)$ and $J_{B_1}(A_0)$ all depend on the fluence as discussed earlier.  $T_c$ initially decreases slightly, even though the channel asymmetry is decreasing, because the overall $T_{K1}$ is decreasing faster. For larger fluences, $J_{A_1}$ increases again, allowing $T_c$ to increase by up to a factor of three as the system is driven through the channel degenerate point. 

\section{Experimental considerations\label{sec:exp}}

For Floquet engineering in real materials, it is important to consider the relevant time and energy scales, and eliminate significant heating effects.  Experimentally, a pulse of length $T_{pulse}$ is applied, with a much shorter Floquet period, $T = 2\pi/\Omega$.  Within this pulse, there must be a pre-thermalization regime where the Floquet-Kondo model is relevant; the existence of this pre-thermalization regime requires the absence of significant heating effects~\cite{Oka_review_2019}. In addition, within this regime, the spins and electrons must relax to the state described by the emergent Floquet-Kondo model, where measurements could be performed to test our predictions.  In the Kondo model far from criticality, this relaxation time is governed by the Kondo temperature, $T_{rel} \sim h/(k_{B}T_{K})$. If polarization averaging in time is being used, the relaxation time must be longer than the polarization time, $T_p$ in order for the spins to feel the average emergent Kondo couplings rather than the time-dependent couplings; $T_p \gtrsim 10 T$ generally ensures effective averaging~\cite{QuitoFlintPRL2021}. As the Floquet-Kondo impurity models approach the quantum critical points given by multi-channel degeneracies, the relaxation time will diverge, and the appropriate behavior after $h/(k_{B}T_{K})$ will be given by a quantum quench to the critical point~\cite{DeGrandi2010,eckstein2017arxiv}. We can find three-channel degeneracies in the case of circular polarization, but four-channel degeneracy does require polarization averaging. It is an open question what polarization averaging does in such a situation, as the model is quantum critical only after averaging; the Kondo couplings are likely disordered in time, and numerical studies are necessary to resolve this question.  For Floquet-Kondo lattices, these quantum critical points are likely hidden by the composite pair superconductivity (or potentially other phases), and so the relaxation time will be determined by those energy scales instead.

The largest emergent couplings are found close to the resonances, for $n\Omega =|\epsilon_f|$, and so $\Omega \sim |\epsilon_f|$, on the order of several electron volts. The relaxation times for Kondo materials are relatively long, as $T_K \approx 100$K gives $T_{rel} \approx 10^{-13}$s.  A long pulse, $T_{pulse}\gtrsim 100$ fs, is therefore required to ensure the spins have fully equilibrated to the new Kondo interactions.  The physics of interest occurs when the dimensionless vector potential $A_{0}\sim1-10$, which corresponds to a laser intensity of $I=c\epsilon_{0}A_{0}^{2}\left|\frac{\Omega\hbar}{ea_{0}}\right|^{2}\sim10^{17}-10^{18}W/m^{2}$ for $\Omega\hbar\sim1eV$ and lattice scales $a_{0}\sim1\lyxmathsym{\AA}$. The simultaneous requirement of a relatively long pulse and large fluence will make these experiments challenging but not impossible.

To have a well-defined transient regime where the Floquet-Kondo effect is realized, the time to thermalize the electrons has to be larger than any other time scale, or practically, the off-resonance condition has to be imposed, avoiding direct absorption of photons by any internal degree of freedom of the system~\cite{McIver_Nature_2020}.  In a simple one-band, infinite-$U$ Anderson model, there will be linear (in time) energy absorption for $0\leq -|\epsilon_f| +\Omega \leq D/2$, where there can be single photon excitations from the occupied local $f$-level to the empty conduction states, with $D$ the total conduction electron bandwidth.  Otherwise, the energy absorption is exponentially suppressed.  We are interested in $\Omega < |\epsilon_f|$, which naturally satisfies this requirement. For a single-band model, there are no direct transitions within the conduction band, although this will not be generically true in materials.  Experimental realizations will also have to avoid frequencies associated with direct transitions (as seen, for example, in the optical conductivity) and from $f-f$ transitions, such as from $J=5/2$ to the excited $J=7/2$ levels in Ce.  Nevertheless, with some care, it is likely possible to achieve the required long thermalization times in Kondo materials.

After the system reaches the Floquet plateau with new couplings, it becomes a matter of measuring features that follow from this reduction in the coupling anisotropy. The enhancement of $T_{c}$ for superconductivity is a first hint that the anisotropy of the Kondo channels has been reduced. For the particular case of composite order of multi-channel systems, we mention the following features that may be experimentally verified~\cite{Flint_NatPhys_2008}: (1)  The symmetry of the superconducting gap follows from the irreps of the two channels, leading to $d$ of $g$-wave superconductors;  (2) the Andreev reflection in tunneling measurements is enhanced; (3) for the normal state, an upturn in the NMR relaxation rate should be observed. Obviously, these measurements are challenging as they have to be done while the laser field is on.

\section{Conclusions\label{sec:Conclusions}} 

We have shown that Floquet engineering provides a powerful new way to tune multi-channel Kondo models, where the number of channels is controlled by a combination of the light polarization and lattice symmetries, with different polarization, intensity, and frequency choices leading to different channel degeneracies. The channel anisotropy, often undesirable, can be significantly reduced or even eliminated by changing the intensity of the field. A tunable driven quantum dot is possibly the simplest experimental realization of our proposal. The toy model and realistic spin-orbit coupled Ce model describe similar physics, showing that this proposal is relevant for Kondo lattice materials. While we explored the square lattice case, our results are generic and can be readily extrapolated to other lattices and f-electron materials. The protocols we presented provide a natural platform to investigate the multicritical behavior of Kondo impurities and Kondo lattices, with significant enhancement of the critical superconducting temperature predicted for composite pair superconductivity. 

The treatment here was within a mean-field theory for the composite order and within the time-independent effective Floquet theory. We leave for future work the study of effects beyond the mean-field theory and how they affect the stability of the Floquet transient plateau.

\subsection*{Acknowledgments} We acknowledge useful discussions with Milan Kornja\v ca, Tha\'{i}s Trevisan, and Bing Li. V.L.Q and R.F. were supported by the U.S. Department of Energy, Office of Science, Basic Energy Sciences, under Award No. DE-SC0015891. VLQ and RF thank the Aspen Center for Physics, supported by the NSF Grant PHY-1607611, for hospitality. 

\bibliographystyle{apsrev4-1}

\appendix

\section{Toy model Kondo couplings~\label{sec:toy_model_App}}

In this Appendix, we calculate the emergent Kondo couplings in detail for the toy model with fixed polarization (circular and linear) and polarization averaging.  We consider both nearest-neighbor (NN) hybridization, $V_0$ and next-nearest-neighbor (NNN) hybridization, $V_1$, in order to capture additional channels not found in the main text, which fixed $V_1 = 0$.

\subsection{Circular polarization}

The Floquet-dressed Wannier conduction electrons $\psi_{\boldsymbol{R}_{j},\sigma}^{\left(n\right)}$ become

\begin{align}
\frac{\psi_{\boldsymbol{R}_{j},\sigma}^{\left(n\right)}}{\mathcal{J}_{n}\left(A_{0}\right)} & =\left(c_{\boldsymbol{R}_{j}+\hat{x},\sigma}+\left(-1\right)^{n}c_{\boldsymbol{R}_{j}-\hat{x},\sigma}\right)+\nonumber \\
 & +e^{\mp in\frac{\pi}{2}}\left(c_{\boldsymbol{R}_{j}+\hat{y},\sigma}+\left(-1\right)^{n}c_{\boldsymbol{R}_{j}-\hat{y},\sigma}\right)+\nonumber \\
 & +\eta^{\left(n\right)}e^{\mp in\frac{\pi}{4}}\left(c_{\boldsymbol{R}_{j}+\hat{d}_{1},\sigma}+\left(-1\right)^{n}c_{\boldsymbol{R}_{j}-\hat{d}_{1},\sigma}\right)+\nonumber \\
 & +\eta^{\left(n\right)}e^{\pm in\frac{\pi}{4}}\left(c_{\boldsymbol{R}_{j}+\hat{d}_{2},\sigma}+\left(-1\right)^{n}c_{\boldsymbol{R}_{j}-\hat{d}_{2},\sigma}\right),
\end{align}
for $\pm = $ LCP/RCP,
where we define the NNN lattice vectors, $\hat{d}_{1,2}=\hat{x}\pm\hat{y}$, and

\begin{equation}
\eta^{\left(n\right)}=\frac{V_{1}\mathcal{J}_{n}\left(\sqrt{2}A_{0}\right)}{V_{0}\mathcal{J}_{n}\left(A_{0}\right)}.\label{eq:eta-factor}
\end{equation}
For convenience, we define a basis of momentum
space form-factors that transform as irreps of $C_{4}$:

\begin{align}
\Phi_{{\bf k}}^{\Gamma_{1a,2a}}& =\cos k_{x}\pm\cos k_{y},\nonumber \\
\Phi_{{\bf k}}^{\Gamma_{3a},\Gamma_{4a}} & =\pm i\sin k_{x}+\sin k_{y},\nonumber \\
\Phi_{{\bf k}}^{\Gamma_{1b}} & =2\cos k_{x}\cos k_{y},\nonumber \\
\Phi_{{\bf k}}^{\Gamma_{2b}} & =2i\sin k_{x}\sin k_{y},\nonumber \\
\Phi_{{\bf k}}^{\Gamma_{3b},\Gamma_{4b}} & =\mp\sqrt{2}(\cos k_{x}\sin k_{y}\pm i\cos k_{y}\sin k_{x}).\label{eq:C4v-toy-basis}
\end{align}
All of these and similar functions in this Appendix are orthonormal, with $\int d\boldsymbol{k}\Phi_{{\bf k}}^{\Upsilon}\Phi_{{\bf k}}^{\Upsilon^{\prime}}=4\pi^{2}\delta_{\Upsilon\Upsilon^{\prime}}$. We label $\Gamma_{i}$ the irreps of the $C_{4}$ group. 

For LCP, the relevant form-factors for each $n$ in Eq.~\eqref{eq:wannier-ff} are
\begin{equation}
\frac{\Phi_{\boldsymbol{k}}^{\left(n\right)}}{\mathcal{J}_{n}\left(A_{0}\right)}=\begin{cases}
\Phi_{{\bf k}}^{\Gamma_{1a}}+\left(-1\right)^{m}\eta^{\left(4m\right)}\Phi_{{\bf k}}^{\Gamma_{1b}}, & n=4m\\
\Phi_{{\bf k}}^{\Gamma_{3a}}+\left(-1\right)^{m}\eta^{\left(4m+1\right)}\Phi_{{\bf k}}^{\Gamma_{3b}}, & n=4m+1\\
\Phi_{{\bf k}}^{\Gamma_{2a}}+\left(-1\right)^{m}\eta^{\left(4m+2\right)}\Phi_{{\bf k}}^{\Gamma_{2b}}, & n=4m+2\\
\Phi_{{\bf k}}^{\Gamma_{4a}}+\left(-1\right)^{m}\eta^{\left(4m+3\right)}\Phi_{{\bf k}}^{\Gamma_{4b}}. & n=4m+3.
\end{cases}
\end{equation}
We omit the spin indices, as the hybridizations are diagonal here. For RCP, indices 3 and 4 are permuted. It becomes clear that, for each $n$, $\Phi_{\boldsymbol{k}}^{\left(n\right)}$ transforms as an irrep of $C_{4}$. Generically, with $V_{1}$ finite, a given $n$ stills lead to a \emph{single} channel, as $V_0$ and $V_1$ terms of $\Phi_{\boldsymbol{k}}^{\left(n\right)}$ transform identically under $C_{4}$ operations. The NN Kondo couplings, found by setting $V_{1}=0$ are given by Eq.~\eqref{eq:toy-circular-couplings}. 

\begin{figure}
\includegraphics[width=0.97\columnwidth]{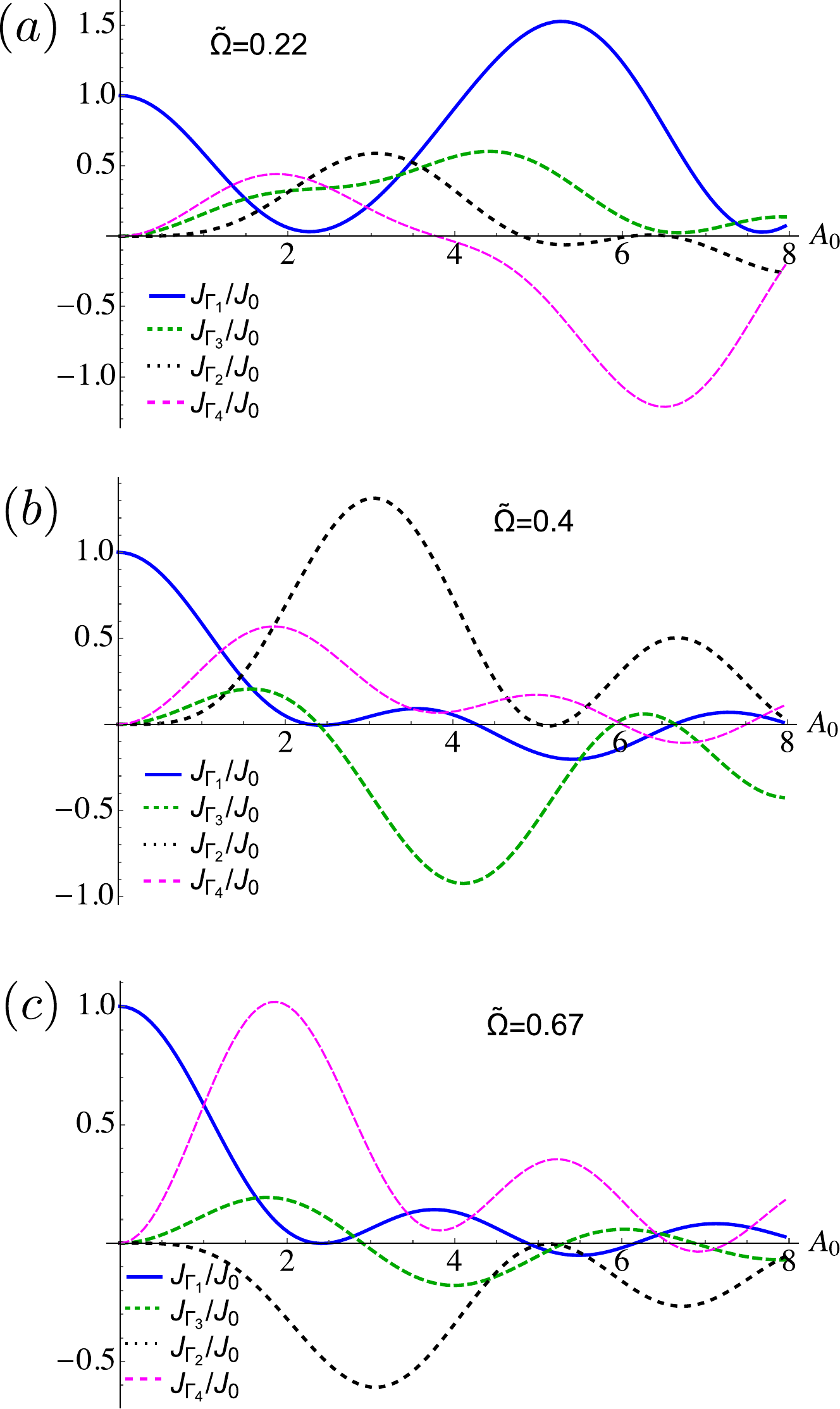}\caption{Toy model coupled to LCP light, for different off-resonance choices of $\tilde{\Omega}=\Omega/\left|\epsilon_{f}\right|$,
keeping only nearest-neighbor hybridizations. Initially, without the Floquet potential, we assume that only the $\Gamma_{1}$ coupling is finite. As $A_{0}$ increases, the $\Gamma_{1}$ channel has a reduced intensity, while some of the $\Gamma_{2,3,4}$
are enhanced. \label{fig:Toy-model-LCP}}
\end{figure}

\subsection{Linear polarization~\label{subsec:Linear-polarization}}

\begin{table}
\begin{centering}
\begin{tabular}{|c|c|}
\hline 
 & Coefficients\tabularnewline
\hline 
\hline 
$\alpha_{A_{1a}}$  & $\frac{1}{2}\left[\mathcal{J}_{n}\left(\sqrt{2}A_{0}\cos\psi\right)+\mathcal{J}_{n}\left(\sqrt{2}A_{0}\sin\psi\right)\right]$\tabularnewline
\hline 
$\alpha_{A_{1b}}$  & $\frac{V_{1}}{2V_{0}}\left[\mathcal{J}_{n}\left(2A_{0}\left(\cos\psi+\sin\psi\right)\right)+\mathcal{J}_{n}\left(2A_{0}\left(\cos\psi-\sin\psi\right)\right)\right]$\tabularnewline
\hline 
$\alpha_{B_{1}}$  & $\frac{1}{2}\left[\mathcal{J}_{n}\left(\sqrt{2}A_{0}\cos\psi\right)-\mathcal{J}_{n}\left(\sqrt{2}A_{0}\sin\psi\right)\right]$\tabularnewline
\hline 
$\alpha_{B_{2}}$  & $\frac{V_{1}}{2V_{0}}\left[\mathcal{J}_{n}\left(2A_{0}\left(\cos\psi+\sin\psi\right)\right)-\mathcal{J}_{n}\left(2A_{0}\left(\cos\psi-\sin\psi\right)\right)\right]$\tabularnewline
\hline 
$\alpha_{E_{x}}$  & $i/\sqrt{2}\mathcal{J}_{n}\left(\sqrt{2}A_{0}\cos\psi\right)$\tabularnewline
\hline 
$\alpha_{E_{y}}$  & $i/\sqrt{2}\mathcal{J}_{n}\left(\sqrt{2}A_{0}\sin\psi\right)$\tabularnewline
\hline 
$\alpha_{E_{c}}$  & $\frac{V_{1}}{2V_{0}}i\left[\mathcal{J}_{n}\left(2A_{0}\left(\cos\psi+\sin\psi\right)\right)+\mathcal{J}_{n}\left(2A_{0}\left(\cos\psi-\sin\psi\right)\right)\right]$\tabularnewline
\hline 
$\alpha_{E_{d}}$  & $\frac{V_{1}}{2V_{0}}i\left[\mathcal{J}_{n}\left(2A_{0}\left(\cos\psi+\sin\psi\right)\right)-\mathcal{J}_{n}\left(2A_{0}\left(\cos\psi-\sin\psi\right)\right)\right]$\tabularnewline
\hline 
\end{tabular}
\par\end{centering}
\caption{Coefficients of a generic linearly polarized light decomposed into
the $g_{\boldsymbol{k}}$ basis {[}Eq.~(\ref{eq:C4v-toy-basis}){]},
combining Eqs.~(\ref{eq:linear-coeff-even})~and~(\ref{eq:linear-coeff-odd}).
For a generic angle $\psi$, the $\alpha_{E_{x}}$ and $\alpha_{E_{y}}$
channels have different coefficients, showing that the symmetry is
broken down to $C_{2v}.$ \label{tab:Coefficients-linear-pol}}
\end{table}

We start with a generic linear polarization, making an angle
$\psi$ with the $x$ axis. Generically, the remaining symmetry group is $C_{2}$, with a few special angles that preserve mirror symmetries and are $C_{2v}$. As we will eventually recover $C_{4v}$ group \emph{after} the polarization average is performed in the
next subsection, we define the functions that transform as irreps under the
$C_{4v}$ group,

\begin{align}
\Phi_{{\bf k}}^{A_{1a}} & =\cos k_{x}+\cos k_{y},\,\,\,\Phi_{{\bf k}}^{A_{1b}}=2\cos k_{x}\cos k_{y}\nonumber \\
\Phi_{{\bf k}}^{B_{1}} & =\cos k_{x}-\cos k_{y},\,\,\,\,\,\Phi_{{\bf k}}^{B_{2}}=2\sin k_{x}\sin k_{y},\nonumber \\
\Phi_{{\bf k}}^{E_{xa}} & =\sqrt{2}\sin k_{x},\,\,\,\,\Phi_{{\bf k}}^{E_{ya}}=\sqrt{2}\sin k_{y},\nonumber \\
\Phi_{{\bf k}}^{E_{xb}}= & 2\cos k_{y}\sin k_{x},\,\,\,\,\,\Phi_{{\bf k}}^{E_{yb}}=2\cos k_{x}\sin k_{y}.\label{eq:g-basis}
\end{align}
Note that we are now using the common notation $A,B,E$ for $C_{4v}$ irreps. 
The functions $A_{1a}$, coming from NN hybridizations,
and $A_{1b}$, coming from NNN hybridizations, transform identically
under $C_{4v}$; we use the same notation to distinguish the $E_a$ and $E_b$ form-factors coming from NN and NNN hybridizations, respectively. We do not calculate the couplings for $A_{2}$ irrep, as those require even further neighbor initial hybridizations; for completeness, these would give $\Phi_{{\bf k}}^{A_{2}}=2\sqrt{2}\sin k_{x}\sin k_{y}\left(\cos k_{x}-\cos k_{y}\right)$.

Now we turn to the Floquet-Wannier functions derived from Eq.~\eqref{eq:psi-Floquet}, where for linear polarization we have the Bessel function argument, $A_{l}=\sqrt{2}|\mathbf{\delta}_{l}|A_{0}\cos\left(\psi-\phi_{l}\right)$
and phase, $\beta_{l}=0$ on link $\mathbf{\delta}_l$. For $n$ even, the resulting form-factor is,
\begin{align}
\Phi_{\boldsymbol{k}}^{\left(n,e\right)}\! & =\mathcal{J}_{n}\!\left(\!\sqrt{2}A_{0}\cos\psi\!\right)\cos k_{x}+\mathcal{J}_{n}\!\left(\!\sqrt{2}A_{0}\sin\psi\!\right)\cos k_{y}\nonumber \\
& +\frac{V_{1}}{V_{0}}\left[\mathcal{J}_{n}\left(2A_{0}\left(\cos\psi+\sin\psi\right)\right)\cos\left(k_{x}+k_{y}\right)\right.\nonumber \\
+ & \left.\mathcal{J}_{n}\left(2A_{0}\left(\cos\psi-\sin\psi\right)\right)\cos\left(k_{x}-k_{y}\right)\right].\label{eq:fn-linear-toy-even}
\end{align}
Rewriting in terms of the $g_{\boldsymbol{k}}$ basis, defined in
Eq.~(\ref{eq:g-basis}),

\begin{align}
\Phi_{\boldsymbol{k}}^{\left(n,e\right)} & =\alpha_{A_{1a}}\Phi_{{\bf k}}^{A_{1a}}+\alpha_{A_{1b}}\Phi_{{\bf k}}^{A_{1b}}+\alpha_{B_{1}}\Phi_{{\bf k}}^{B_{1}}+\alpha_{B_{2}}\Phi_{{\bf k}}^{B_{2}},\label{eq:fn-linear-toy-decomp}
\end{align}
with the coefficients

\begin{align}
\alpha_{A_{1a},B_{1}} & =\frac{1}{2}\left[\mathcal{J}_{n}\left(\sqrt{2}A_{0}\cos\psi\right)\pm\mathcal{J}_{n}\left(\sqrt{2}A_{0}\sin\psi\right)\right],\nonumber \\
\alpha_{A_{1b},B_{2}} & =\frac{V_{1}}{2V_{0}}\left[\mathcal{J}_{n}\left(2A_{0}\left(\cos\psi+\sin\psi\right)\right)\pm\right.\nonumber \\
 & \pm\left.\mathcal{J}_{n}\left(2A_{0}\left(\cos\psi-\sin\psi\right)\right)\right]\label{eq:linear-coeff-even}
\end{align}
This decomposition will be particularly useful when we consider the polarization average. For now, the $n$-even hybridizations lead to a single channel.

For $n$ odd, the cosines are replaced by sines, as

\begin{align}
\Phi_{\boldsymbol{k}}^{\left(n,o\right)}\! & =i\mathcal{J}_{n}\!\left(\!\sqrt{2}A_{0}\cos\psi\!\right)\sin k_{x}+i\mathcal{J}_{n}\!\left(\!\sqrt{2}A_{0}\sin\psi\! \right)\sin k_y \nonumber \\ & +i\frac{V_{1}}{V_{0}}\left[\mathcal{J}_{n}\left(2A_{0}\left(\cos\psi+\sin\psi\right)\right)\sin\left(k_{x}+k_{y}\right)\right.\nonumber \\
+ & \left.\mathcal{J}_{n}\left(2A_{0}\left(\cos\psi-\sin\psi\right)\right)\sin\left(k_{x}-k_{y}\right)\right]\label{eq:fn-linear-toy-odd}
\end{align}
Decomposing again in the $g_{\boldsymbol{k}}$ basis, 
\begin{align}
\Phi_{\boldsymbol{k}}^{\left(n,o\right)} & =\alpha_{E_{xa}}\Phi_{{\bf k}}^{E_{xa}}+\alpha_{E_{ya}}\Phi_{{\bf k}}^{E_{ya}}+\alpha_{E_{xb}}\Phi_{{\bf k}}^{E_{xb}}+\alpha_{E_{yb}}\Phi_{{\bf k}}^{E_{yb}},\label{eq:fn-linear-toy-decomp-2}
\end{align}
with

\begin{align}
\alpha_{E_{xa}} & =\frac{i}{\sqrt{2}}\mathcal{J}_{n}\left(\sqrt{2}A_{0}\cos\psi\right),\nonumber \\
\alpha_{E_{ya}} & =\frac{i}{\sqrt{2}}\mathcal{J}_{n}\left(\sqrt{2}A_{0}\sin\psi\right),\nonumber \\
\alpha_{E_{(x,y)b}} & =\frac{V_{1}}{2V_{0}}i\left[\mathcal{J}_{n}\left(2A_{0}\left(\cos\psi+\sin\psi\right)\right)\right.\nonumber \\
 & \left.\mp\mathcal{J}_{n}\left(2A_{0}\left(\cos\psi-\sin\psi\right)\right)\right].\label{eq:linear-coeff-odd}
\end{align}

The $n-$odd hybridizations lead also to a single channel and, therefore, for linearly polarized light, we have two Kondo channels. The collection of all the $\alpha_{\Upsilon}$ coefficients is listed in Table~\ref{tab:Coefficients-linear-pol}.

\subsection{Quasi-monochromatic light~\label{subsec:quasi-monochromatic}}

In this subsection, we detail the averaging
over ensembles of monochromatic light. While the average can be performed over any path on the surface of the Poincar\'e sphere, we focus on two cases here: the LCP/RCP average, which is the simplest; and the average over an ensemble of linearly polarized light, which is more straightforward to generate experimentally.

\subsubsection{LCP/RCP average}

In order to perform the average over LCP/RCP, we go back to Eq.~\eqref{eq:toy-circular-couplings}.
We can anticipate that the chiral channels $\Gamma_{3}$ and $\Gamma_{4}$ are the ones that will change the most, as chirality is eliminated under the averaging procedure. The average restores the $x\rightleftarrows y$ symmetry. By considering NN hybridizations, for simplicity, we find the $\Gamma_3$ average,

\begin{align}
\left\langle f_{\boldsymbol{k}}^{\Gamma_{3}}\left(f_{\boldsymbol{k}^{\prime}}^{\Gamma_{3}}\right)^{*}\right\rangle _{\text{LCP/RCP}}=2\left[\sin k_{x}\sin k_{x}^{\prime}+\sin k_{y}\sin k_{y}^{\prime}\right].
\end{align}

The average leads to two channels that are \emph{intrinsically} \emph{degenerate}, as required by the underlying $C_{4v}$ symmetry. In fact, $k_{x}$
and $k_{y}$ appear symmetrically, implying that this corresponds to the two-dimensional irrep of $C_{4v}$ ($E$). Proceeding similarly for $\Gamma_{4}$, we see that

\begin{equation}
\left\langle f_{\boldsymbol{k}}^{\Gamma_{4}}\left(f_{\boldsymbol{k}^{\prime}}^{\Gamma_{4}}\right)^{*}\right\rangle_{\text{LCP/RCP}} =\left\langle f_{\boldsymbol{k}}^{\Gamma_{3}}\left(f_{\boldsymbol{k}^{\prime}}^{\Gamma_{3}}\right)^{*}\right\rangle_{\text{LCP/RCP}} ,
\end{equation}
showing that both chiral channels lead to the same degenerate $p_{x}/p_{y}$
channels.

For simplicity, we consider only the leading order terms of the hybridizations for each channel. The overall Kondo couplings after the LCP/RCP average are 

\begin{align}
J_{A_{1}} & =J_{0}\sum_{m=-\infty}^{\infty}\frac{\mathcal{J}_{4m}^{2}\left(A_{0}\right)}{1+4m\tilde{\Omega}},\nonumber \\
J_{E} & =J_{0}\sum_{m=-\infty}^{\infty}\frac{\mathcal{J}_{2m+1}^{2}\left(A_{0}\right)}{1+\left(2m+1\right)\tilde{\Omega}},\nonumber \\
J_{B_{2}} & =J_{0}\sum_{m=-\infty}^{\infty}\frac{\mathcal{J}_{4m+2}^{2}\left(A_{0}\right)}{1+\left(4m+2\right)\tilde{\Omega}}.\label{eq:toy-LCP-RCP}
\end{align}
Here, $J_{E}$ corresponds to the two degenerate channels $E_{x}$
and $E_{y}$. Remember that this channels only include the NN hybridizations, and further neighbor couples will generate $J_{B_1}$. 

\begin{figure}
\includegraphics[width=0.8\columnwidth]{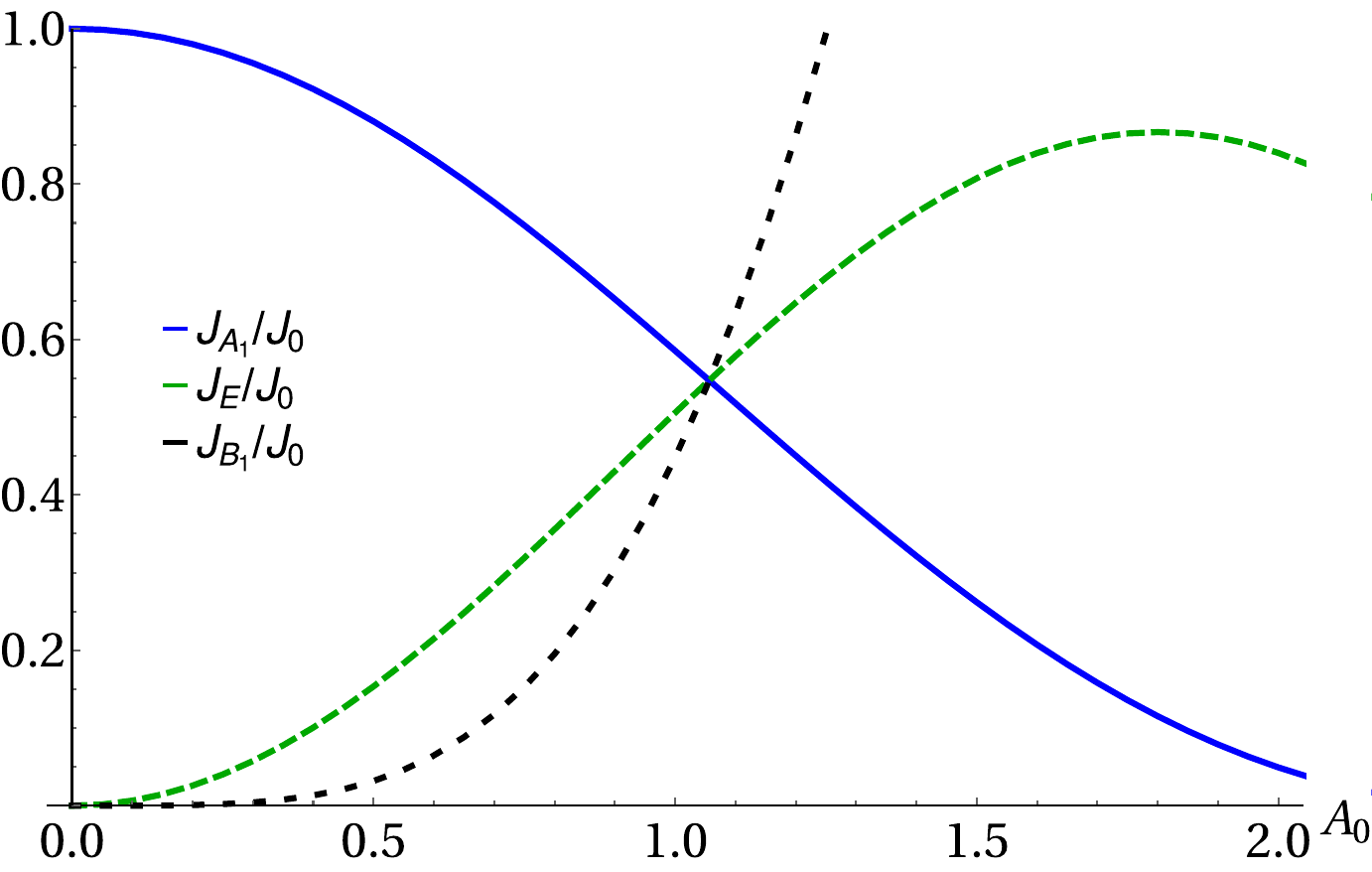}\caption{Possibility of realizing a four-channel symmetric Kondo model starting with a single channel. For $A_a\approx1.05$ and $\Omega\approx0.485$, the four channels are degenerate. While this choice of $\tilde{\Omega}=\Omega/\left|\epsilon_{f}\right|$ is potentially close to the 2-photon resonance, the 4-channel degenerate cases are generically possible, as the two $E$ channels are intrinsically degenerate, and the two other channels, $s$ and $d$ wave channels can be manipulated by two parameters, the frequency and the amplitude of the radiation.  \label{fig:Toy-model-LCP-4channels}}
\end{figure}

By changing the fluence and frequency, it is possible to tune to a point where all four channels, $A_1$, $B_1$, and  $E$ are degenerate. This is shown in Fig.~\ref{fig:Toy-model-LCP-4channels}. For the model considered here, this point is close to the 2-photon resonance, but the possibility of tuning the amplitude of the $A_1$ and $B_1$ channels by varying frequency and intensity is generic to any model.

\subsubsection{Type II Glauber light}

We now address the type II Glauber light, that yields the same channels as the LCP/RCP, as required by symmetry, but with different relative strengths. We use the explicit decomposition of Eq.~(\ref{eq:g-basis}), and the coefficients are defined in Table~\ref{tab:Coefficients-linear-pol}.
By symmetry, it follows that $\left\langle \left|\alpha_{E_{xa}}\right|^{2}\right\rangle =\left\langle \left|\alpha_{E_{ya}}\right|^{2}\right\rangle $
and $\left\langle \left|\alpha_{E_{xb}}\right|^{2}\right\rangle =\left\langle \left|\alpha_{E_{yb}}\right|^{2}\right\rangle $.
Also, it follows that $\left\langle \alpha_{X}\alpha_{Y}\right\rangle \propto\delta_{X,Y}$,
with $X,Y$ labeling any $C_{4v}$ irreps -- that is, crossed terms that
connect different irreps always vanish. These conditions are, in fact, \emph{necessary}
for the emergence of the two-dimensional irrep $E$.  $E_{a,b}$ denote different versions of the same irreps, corresponding to NN and NNN hybridizations, but all cross terms of the type $E_{xa}$ and $E_{xb}$ vanish as well.

Collecting these results, we find a collection of  effective Kondo channels,
\begin{align}
J_{A_{1a}} & =J_{0}\sum_{m=-\infty}^{\infty}\frac{\left\langle \left|\alpha_{A_{1a}}\right|^{2}\right\rangle }{1+2m\tilde{\Omega}},\nonumber \\
J_{A_{1b}} & =J_{0}\sum_{m=-\infty}^{\infty}\frac{\left\langle \left|\alpha_{A_{1b}}\right|^{2}\right\rangle}{1+2m\tilde{\Omega}},\nonumber \\
J_{B_{1}} & =J_{0}\sum_{m=-\infty}^{\infty}\frac{\left\langle \left|\alpha_{B_{1}}\right|^{2}\right\rangle }{1+2m\tilde{\Omega}},\nonumber \\
J_{B_{2}} & =J_{0}\sum_{m=-\infty}^{\infty}\frac{\left\langle \left|\alpha_{B_{2}}\right|^{2}\right\rangle }{1+2m\tilde{\Omega}},\nonumber \\
J_{E_{xa,ya}} & =J_{0}\sum_{m=-\infty}^{\infty}\frac{\left\langle \left|\alpha_{E_{xa}}\right|^{2}\right\rangle }{1+\left(2m+1\right)\tilde{\Omega}},\nonumber \\
J_{E_{xb,yb}} & =J_{0}\sum_{m=-\infty}^{\infty}\frac{\left\langle \left|\alpha_{E_{xb}}\right|^{2}\right\rangle  }{1+\left(2m+1\right)\tilde{\Omega}}.
\end{align}
Here, we separate the Kondo couplings by their form-factors, and the dominant channels are $A_{1a},B_1,E_{xa,ya}$, that come solely from nearest-neighbor hybridizations. The channels $A_{1b},B_2,E_{xb,yb}$ are proportional to $(V_1/V_0)^2$ and are included here from completeness, but neglected in the main text; $a$ and $b$ couplings will have different form factors and strengths, but will not lead to different channels, as they are not orthogonal at the $\Gamma$ point.  For the channel $E$, the $k_{x}\rightleftarrows k_{y}$ symmetry is again a consequence of the two-dimensional character of the irrep. 

We end by noting that the arguments and derivation above hold for any combination of linearly polarized light that keeps the $C_{4v}$ symmetry. Other possibilities include, for example, the simple average of horizontal and vertical linear polarization in the case of quantum dots or if further neighbor hybridizations can be neglected in the lattice case.

\subsection{Starting with multiple equilibrium channels}

All the derivations above assumed a single non-zero ($s$-wave) hybridization profile. In what follows, we address the case with two initial, non-degenerate channels for valence fluctuations, mediated by
conduction electrons with $s$ and $d_{x^2-y^2}$ wave symmetries, respectively. We illustrate this effect for the type II Glauber light, and other cases follow analogously. The first step is to perform the Floquet dressing of the initial $d_{x^2-y^2}$, called $f_d$. For $n$ even, we get for NN hybridization, 

\begin{align}
f_{d}^{\left(n\,\text{even}\right)}\left(\boldsymbol{k}\right) & =\mathcal{J}_{n}\left(\sqrt{2}A_{0}\cos\psi\right)\cos k_{x}\nonumber \\
 & -\mathcal{J}_{n}\left(\sqrt{2}A_{0}\sin\psi\right)\times\cos k_{y},\label{eq:fn-linear-toy-even-1}
\end{align}
while for $n$ odd

\begin{align}
f_{d}^{\left(n\,\text{odd}\right)}\left(\boldsymbol{k}\right) & =i\mathcal{J}_{n}\left(\sqrt{2}A_{0}\cos\psi\right)\sin k_{x}\nonumber \\
 & -i\mathcal{J}_{n}\left(\sqrt{2}A_{0}\sin\psi\right)\sin k_{y}.\label{eq:fn-linear-toy-odd-1}
\end{align}
The important difference as compared to the dressed $s$-wave hybridizations is the relative minus signs between the two lines of each equation. We assume that these valence fluctuations involve an excited state of energy $a\left|\epsilon_{f}\right|$, with $a>0$. By performing the same decomposition as in the last subsection, we find that each effective Kondo couplings now has contributions from both equilibrium channels, with different $n$,

\begin{align}
\frac{J_{A_{1}}}{J_0} & =\sum_{m=-\infty}^{\infty}\frac{\left\langle \left|\alpha_{A_{1a}}\right|^{2}\right\rangle}{1+2m\tilde{\Omega}}+\frac{1}{a}\sum_{m=-\infty}^{\infty}\frac{\left\langle \left|\alpha_{B_{1}}\right|^{2}\right\rangle  }{1+2m\left(\tilde{\Omega}/a\right)},\nonumber \\
\frac{J_{B_{1}}}{J_0} & =\sum_{m=-\infty}^{\infty}\frac{\left\langle \left|\alpha_{B_{1}}\right|^{2}\right\rangle  }{1+2m\tilde{\Omega}}+\frac{1}{a}\sum_{m=-\infty}^{\infty}\frac{\left\langle \left|\alpha_{A_{1a}}\right|^{2}\right\rangle  }{1+2m\left(\tilde{\Omega}/a\right)},\nonumber \\
\frac{J_{E_{(x,y)a}}}{J_0} & =\left(1+\frac{1}{a}\right)\left[\sum_{m=-\infty}^{\infty}\frac{\left\langle \left|\alpha_{E_{xa}}\right|^{2}\right\rangle}{1+\left(2m+1\right)\tilde{\Omega}}\right].
\end{align}
Notice that $\left\langle \left|\alpha_{A_{1a}}\right|^{2}\right\rangle=1$ when $A_{0}=0$. For Figure~\ref{fig:Coupling-values}(b)
of the main text, we explicitly set $a=2$.

\section{$\text{Ce}$ model~\label{sec:Ce_model_App}}

In this Appendix, we give the explicit form-factors for the $J=5/2$ Ce model, where the $f$-electrons hybridize with $s$-wave conduction electrons on neighboring sites. As mentioned in the main text, the $\Gamma_{7}$ multiplet 
as the Ce ground state does not lead to hybridizations along the $\pm z$ directions, either in or out of equilibrium. For this reason, we believe this two-dimensional model will give a good approximation for the overall physics of a slab of Ce-based heavy fermion material. We will show in detail that the CP
and LP polarization choices break the
model symmetry down to $C_{4}$ and $C_{2}$, respectively. Again,
by performing the symmetry restoring polarization averages, the model recovers $C_{4v}$ symmetry. 

For generic polarization, and considering both nearest and next-nearest
neighbors, the Wannier functions $\psi_{j,\sigma}^{\left(n\right)}$
at site $j$ and Floquet sector $n$ are written as 

\begin{widetext}

\begin{align}
\psi_{j,\sigma}^{\left(n\right)} & =\sum_{\sigma^{\prime}}\mathcal{J}_{n}\left(A_{x}\right)e^{in\beta\left(0\right)}\left(\begin{array}{cc}
0 & 1\\
1 & 0
\end{array}\right)_{\sigma\sigma^{\prime}}\left(c_{j+\hat{x},\sigma^{\prime}}-\left(-1\right)^{n}c_{j-\hat{x},\sigma^{\prime}}\right)+\mathcal{J}_{n}\left(A_{y}\right)e^{in\beta\left(\frac{\pi}{2}\right)}\left(\begin{array}{cc}
0 & e^{i\pi/2}\\
e^{-i\pi/2} & 0
\end{array}\right)_{\sigma\sigma^{\prime}}\times\nonumber \\
 & \times\left(c_{j+\hat{y},\sigma^{\prime}}-\left(-1\right)^{n}c_{j-\hat{y},\sigma^{\prime}}\right)+\frac{V_{1}}{V_{0}}\left[\mathcal{J}_{n}\left(A_{d_{1}}\right)e^{in\beta\left(\frac{\pi}{4}\right)}\left(\begin{array}{cc}
0 & e^{-3\pi i/4}\\
e^{3\pi i/4} & 0
\end{array}\right)_{\sigma\sigma^{\prime}}\left(c_{j+\hat{d}_{1},\sigma^{\prime}}-\left(-1\right)^{n}c_{j-\hat{d}_{1},\sigma^{\prime}}\right)+\right.\nonumber \\
 & +\left.\mathcal{J}_{n}\left(A_{d_{2}}\right)e^{in\beta\left(-\frac{\pi}{4}\right)}\left(\begin{array}{cc}
0 & e^{3\pi i/4}\\
e^{-3\pi i/4} & 0
\end{array}\right)_{\sigma\sigma^{\prime}}\left(c_{j+\hat{d}_{2},\sigma^{\prime}}-\left(-1\right)^{n}c_{j-\hat{d}_{2},\sigma^{\prime}}\right)\right].\label{eq:hyb_Ce_generic}
\end{align}

\end{widetext}
Here, $A_{l} = A_{ij}$ from Eq.~\eqref{eq:A_l} on a given type of link, while $\beta(\phi_{ij})$ is from Eq.~\eqref{eq:tan-bl}, with $\phi_{ij}$ denoting the angles of the relevant links. The strong spin-orbit coupling causes these hybridization to be off-diagonal in $\sigma \sigma'$, as seen in the Pauli matrix structure.  This form contains all of the necessary information to calculate the Kondo couplings for any polarization choice.

\subsection{Circular polarization and RCP/LCP average}

\begin{figure}
\includegraphics[width=1\columnwidth]{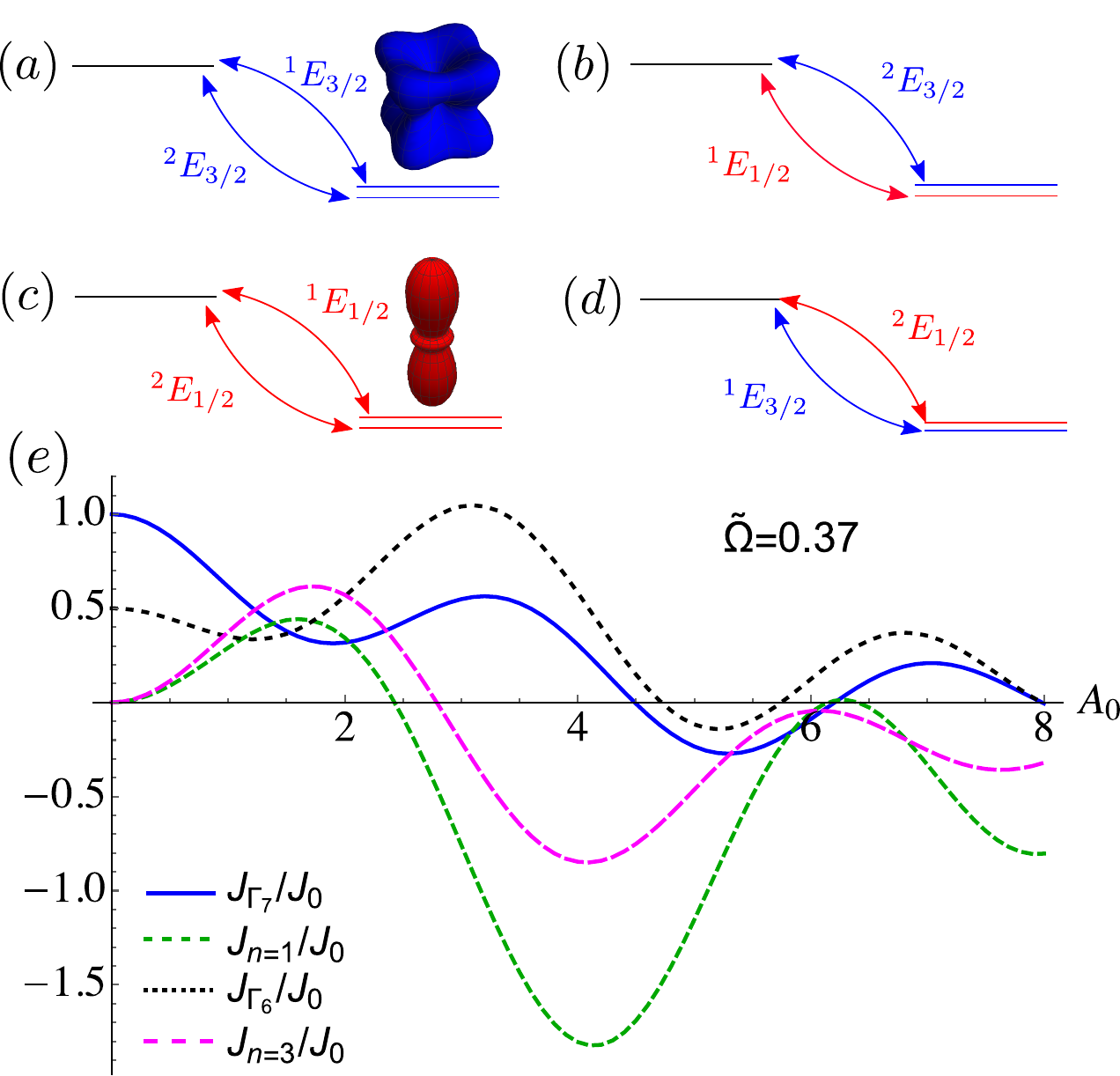}
\caption{In (a)-(d), the four Floquet classes of hybridization profiles for Ce in presence of circularly polarized light. The group symmetry is reduced from $C_{4v}$ to $C_{4}$, as the light breaks time-reversal and inversion symmetries.
The ground-state doublet is degenerate even in the presence of a time-reversal symmetry breaking Floquet field. In (a) and (c), the sectors already present without the Floquet field. The valence fluctuations are mediated by conduction electrons that transform as the two components of the $E_{3/2}$ irrep of $C_{4v}$. In (c), another sector in which the CP light does not reduce the symmetry. The doublet is connected to the excited states by the components of the $E_{1/2}$ irrep of $C_{4v}$. In (b) and (d), valence fluctuation are mediated by conduction electrons that mix the components of $C_{4v}$ irreps. These are, in fact, one-dimensional irreps of the $C_{4}$ group. In (e), the relative strength of the Kondo couplings in terms of the dimensionless vector potential amplitude $A_{0}$ for the Ce model coupled to LCP light for $\tilde{\Omega}=\Omega/\left|\epsilon_{f}\right|=0.37$. \label{fig:Ce-CP}}
\end{figure}

We first consider the circularly polarized case to explore the similarities with the toy model case. For simplicity, we choose $V_1 = 0$. The hybridizations for LCP are found from
Eq.~(\ref{eq:hyb_Ce_generic}) by setting $A_{x}=A_{y}=A_{0}$ and
$\beta=\phi$. There are four distinct Floquet sectors with
momentum space form factors,

\begin{align}
\Phi_{LCP}^{\left(4m\right)}\left(\boldsymbol{k}\right)= & \left(\begin{array}{cc}
0 & i\sin k_{x}-\sin k_{y}\\
i\sin k_{x}+\sin k_{y} & 0
\end{array}\right),\nonumber \\
\Phi_{LCP}^{\left(4m+1\right)}\left(\boldsymbol{k}\right)= & \left(\begin{array}{cc}
0 & \cos k_{x}-\cos k_{y}\\
\cos k_{x}+\cos k_{y} & 0
\end{array}\right),\nonumber \\
\Phi_{LCP}^{\left(4m+2\right)}\left(\boldsymbol{k}\right)= & \left(\begin{array}{cc}
0 & i\sin k_{x}+\sin k_{y}\\
i\sin k_{x}-\sin k_{y} & 0
\end{array}\right),\nonumber \\
\Phi_{LCP}^{\left(4m+3\right)}\left(\boldsymbol{k}\right)= & \left(\begin{array}{cc}
0 & \cos k_{x}+\cos k_{y}\\
\cos k_{x}-\cos k_{y} & 0
\end{array}\right).\label{eq:phi-Ce-model}
\end{align}
The Kondo term for a given Floquet sector will involve the combination,

\begin{equation}
\boldsymbol{M}^{n}\left(\boldsymbol{k},\boldsymbol{k}^{\prime}\right)=\left[\Phi_{LCP}^{n}\left(\boldsymbol{k}\right)\right]^{\dagger}\boldsymbol{\sigma}\Phi_{LCP}^{n}\left(\boldsymbol{k}^{\prime}\right).
\end{equation}
The $n$-even channels present hybridization matrices $\Phi_{\boldsymbol{k}}^{\left(n\right)}$ that are invariant under $C_{4v}$, as $n=4m$ is of type $\Phi_{7\boldsymbol{k}}^{A}$
and $n=4m+2$ is $\Phi_{6\boldsymbol{k}}^{A}$ {[}see Eq.~(\ref{eq:C4v-matrices}) and Fig.~\ref{fig:Ce-CP}(a-c){]} .
The cases of $n=4m+1$ and $n=4m+3$, Fig.~\ref{fig:Ce-CP}~(b) and (d) are unique to Floquet driving. For the $n=4m+1$ sector, for instance, conduction electrons of symmetry $^{2}E_{3/2}$ and $^{1}E_{1/2}$ mediate the $f$-electron fluctuations to the excited state, see Fig.~\ref{fig:Ce-CP}(b). Similar processes could potentially be found without Floquet engineering in the presence of strong magnetic fields, where a doubly degenerate ground state can be restored via accidental degeneracies.
In Fig.~\ref{fig:Ce-CP}(d), we show how the four LCP Kondo couplings evolve with increasing fluence, $A_0$, choosing $\tilde{\Omega}=0.37$.

Next we consider the polarization average over LCP/RCP, which is done at the level of the Kondo Hamiltonian.
By comparing with Eq.~(\ref{eq:phi-Ce-model}), we notice that the $n=4m$ and $n=4m+2$ sectors are not chiral, and are already invariant under $C_{4v}$. These averages do not change $\boldsymbol{M}^{n}\left(\boldsymbol{k},\boldsymbol{k}^{\prime}\right)$
and we have 
\begin{align}
n=4m & \implies\left\langle \boldsymbol{M}^{n}\left(\boldsymbol{k},\boldsymbol{k}^{\prime}\right)\right\rangle =\left(\Phi_{7,\boldsymbol{k}}^{A}\right)^{\dagger}\boldsymbol{\sigma}\Phi_{7,\boldsymbol{k}^{\prime}}^{A}\nonumber \\
n=4m+2 & \implies\left\langle \boldsymbol{M}^{n}\left(\boldsymbol{k},\boldsymbol{k}^{\prime}\right)\right\rangle =\left(\Phi_{6,\boldsymbol{k}}^{A}\right)^{\dagger}\boldsymbol{\sigma}\Phi_{6,\boldsymbol{k}^{\prime}}^{A}\label{eq:n_2_Ce}
\end{align}
The $n$-odd sectors have their $C_{4v}$ symmetry restored after averaging. Just as for the toy model, the averaged $n=4m+1$ and $n=4m+3$ sectors lead to identical terms as these sectors are flipped when the light polarization is reversed. Simple algebra allows to us to re-decompose the averaged
result in terms of the $C_{4v}$-invariant hybridization matrices,
Eq.~(\ref{eq:C4v-matrices}),
\begin{align}
 & \left\langle \boldsymbol{M}^{n=\left(1,3\right)}\left(\boldsymbol{k},\boldsymbol{k}^{\prime}\right)\right\rangle \nonumber \\
 & =\begin{cases}
\frac{1}{2}\left[\left(\Phi_{6,\boldsymbol{k}}^{B}\right)^{\dagger}\sigma_{i}\Phi_{7,\boldsymbol{k}^{\prime}}^{B}+\left(\Phi_{7,\boldsymbol{k}}^{B}\right)^{\dagger}\sigma_{i}\Phi_{6,\boldsymbol{k}^{\prime}}^{B}\right] & ,i=x,y\\
\frac{1}{2}\left[\left(\Phi_{6,\boldsymbol{k}}^{B}\right)^{\dagger}\sigma_{i}\Phi_{6,\boldsymbol{k}^{\prime}}^{B}+\left(\Phi_{7,\boldsymbol{k}}^{B}\right)^{\dagger}\sigma_{i}\Phi_{7,\boldsymbol{k}^{\prime}}^{B}\right] & ,i=z
\end{cases}
\end{align}
These results present some interesting features. First, combined with Eq.~(\ref{eq:n_2_Ce}), it is clear that the Floquet potential can generate more than one \emph{orthogonal} hybridization of the same $C_{4v}$
irrep; here, they are labeled as $\Phi_{\Upsilon}^{A,B}$, for $\Upsilon = 6,7$. These multiple irreps arise because the Floquet field transfers angular momentum to the system, leading to $J=3/2$ and $J=5/2$ versions of the irreps (much like $(p_x, p_y)$ and $(d_{xz}, d_{yz})$ are both $e_g$ irreps in tetragonal symmetry).  This situation is different from $a,b$ irreps in the toy model case, where those irreps are not orthogonal at the $\Gamma$ point. Second, while the $z$ component is diagonal in the channel index, the $x$ and $y$ components are not, leading to a non-$SU(2)$ symmetric interaction. The $x$ and $y$ components combine $\Gamma_{7}$ and $\Gamma_{6}$ in a symmetric way, preserving the underlying $C_{4v}$ symmetry. 

\begin{figure*}
\begin{centering}
\includegraphics[width=2\columnwidth]{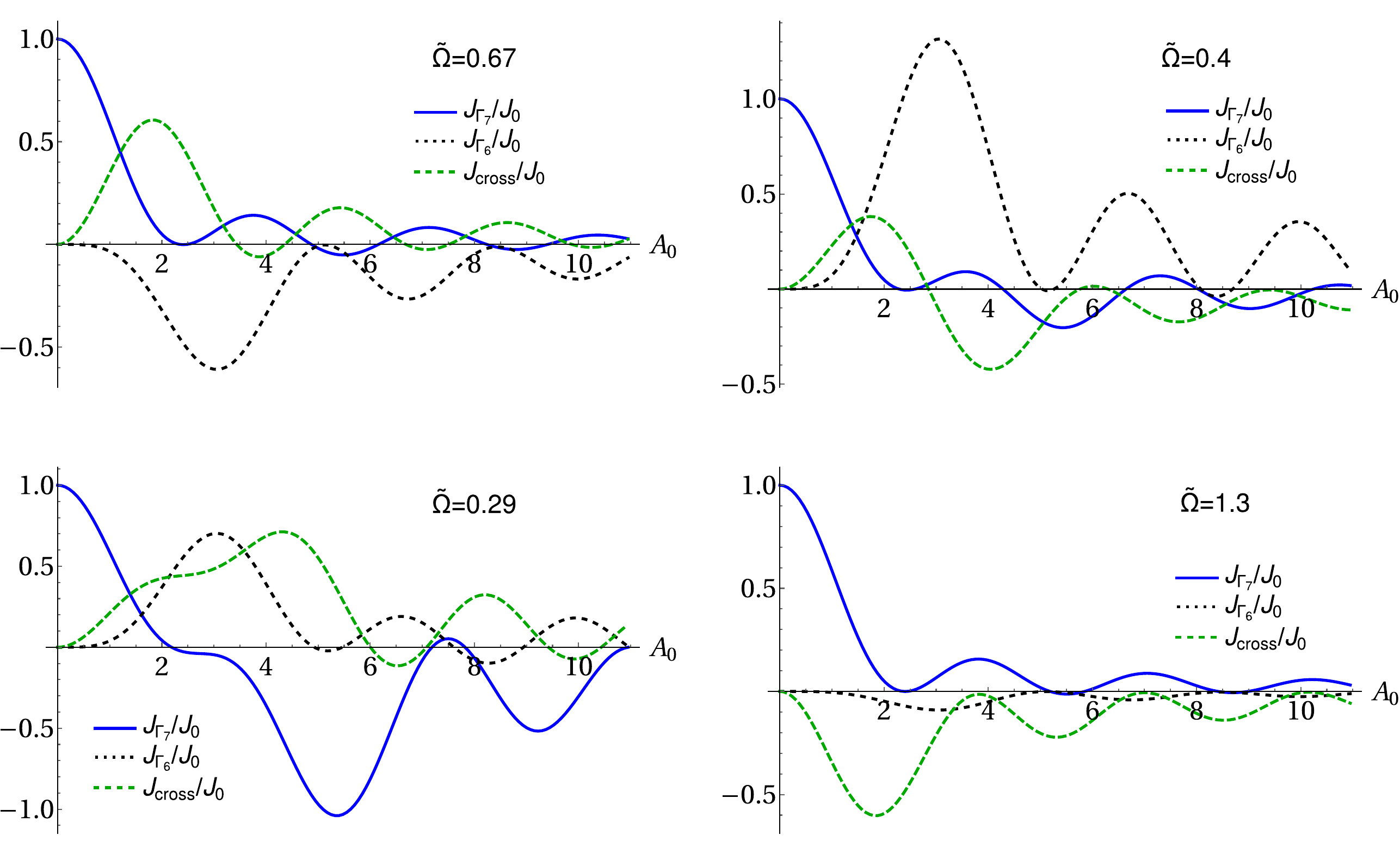}
\par\end{centering}
\caption{The relative strength of the Kondo couplings in terms of the dimensionless vector potential amplitude $A_{0}$ for the Ce model after averaging over LCP and RCP light for a series of choice of $\tilde{\Omega}=\Omega/\left|\epsilon_{f}\right|$. We consider the initial case with only the $\Gamma_{7A}$ channel and call $J_{\text{cross}}$ the terms that mix the $\Gamma_{6B}$ and $\Gamma_{7B}$ irreps. \label{fig:Ce-CP-avg} }
\end{figure*}

In Fig.~\ref{fig:Ce-CP-avg}, we show the relative strengths of the different couplings,
as functions of the dimensionless vector potential $A_{0}$ for several
choices of $\tilde{\Omega}=\Omega/\left|\epsilon_{f}\right|$, $\tilde{\Omega}=0.67,\,0.4,\,0.29$,
and $\tilde{\Omega}=1.3$.

\subsection{Linear polarization}

In this subsection, we explore LP light along an arbitrary direction with initial nearest-neighbor hybridizations. There are now two classes: $n$ even and odd. These are found from Eq.~(\ref{eq:hyb_Ce_generic}) by setting the phase $\beta=0$,

\begin{align}
\Phi_{RCP}^{\left(n\,\text{even}\right)}\left(\boldsymbol{k}\right) & =V_{0}\sigma^{+}\left[i\mathcal{J}_{n}\left(A_{x}\right)\sin k_{x}-\mathcal{J}_{n}\left(A_{y}\right)\sin k_{y}\right]\nonumber \\
 & +V_{0}\sigma^{-}\left[i\mathcal{J}_{n}\left(A_{x}\right)\sin k_{x}+\mathcal{J}_{n}\left(A_{y}\right)\sin k_{y}\right],\\
\Phi_{RCP}^{\left(n\,\text{odd}\right)}\left(\boldsymbol{k}\right) & =V_{0}\sigma^{+}\left[\mathcal{J}_{n}\left(A_{x}\right)\cos k_{x}+i\mathcal{J}_{n}\left(A_{y}\right)\cos k_{y}\right]\nonumber \\
 & +V_{0}\sigma^{-}\left[\mathcal{J}_{n}\left(A_{x}\right)\cos k_{x}-i\mathcal{J}_{n}\left(A_{y}\right)\cos k_{y}\right].
\end{align}
These hybridizations preserve the remaining $C_{2}$ symmetry. Both
$n$ even and odd transform as the $E_{1/2}$ irrep of $C_{2}$,
as expected since this is the only spinorial irrep of the group. 

We now perform the polarization average over an ensemble of LP light.
All the angles $\psi$ are averaged with equal weight, while $\chi$
is kept zero.  We first examine even $n$, where post-averaging, we find the form-factor combinations,
\begin{align}
\left\langle M_{x,y}^{n}\left(\boldsymbol{k},\boldsymbol{k}^{\prime}\right)\right\rangle  & =\left(\begin{array}{cc}
0 & G_{n}\left(\boldsymbol{k},\boldsymbol{k}^{\prime}\right)\\
G_{n}^{*}\left(\boldsymbol{k},\boldsymbol{k}^{\prime}\right) & 0
\end{array}\right),\nonumber \\
\left\langle M_{z}^{n}\left(\boldsymbol{k},\boldsymbol{k}^{\prime}\right)\right\rangle  & =\left(\begin{array}{cc}
H_{n}\left(\boldsymbol{k},\boldsymbol{k}^{\prime}\right) & 0\\
0 & -H_{n}\left(\boldsymbol{k},\boldsymbol{k}^{\prime}\right)
\end{array}\right),
\end{align}
with the functions $H_{n}$ and $G_{n}$ are defined as 
\begin{align}
G_{n}\left(\boldsymbol{k},\boldsymbol{k}^{\prime}\right) & =A_{n}\left(sk_{x}sk_{x}^{\prime}-sk_{y}sk_{y}^{\prime}\right)+\nonumber \\
 & +B_{n}i\left(sk_{x}sk_{y}^{\prime}+sk_{y}sk_{x}^{\prime}\right),\nonumber \\
H_{n}\left(\boldsymbol{k},\boldsymbol{k}^{\prime}\right) & =-A_{n}\left(sk_{x}sk_{x}^{\prime}+sk_{y}sk_{y}^{\prime}\right)+\nonumber \\
 & +B_{n}i\left(sk_{x}sk_{y}^{\prime}-sk_{y}sk_{x}^{\prime}\right),
\end{align}
where we define $sk_i\equiv\sin k_i$.
The coefficients $A_{n}$ and $B_{n}$ are the averages of Bessel functions over $\psi$, 
\begin{align}
A_{n} & =4\times\frac{1}{2\pi}\int_{0}^{2\pi}d\psi\mathcal{J}_{n}^{2}\left(\sqrt{2}A_{0}\cos\psi\right),\nonumber \\
B_{n} & =4\times\frac{1}{2\pi}\int_{0}^{2\pi}d\psi\mathcal{J}_{n}\left(\sqrt{2}A_{0}\cos\psi\right)\mathcal{J}_{n}\left(\sqrt{2}A_{0}\sin\psi\right).
\end{align}
We can then re-decompose these hybridizations in terms of $\Gamma_{6,7}^A$ form factors, as above.
In summary, for linear polarization, even $n$,
\begin{align}
\left.\left\langle \boldsymbol{M}^{n\,\text{even}}\left(\boldsymbol{k},\boldsymbol{k}^{\prime}\right)\right\rangle \right|_{\text{LP}} & =\frac{\left(A_{n}+B_{n}\right)}{2}\left(\Phi_{7,\boldsymbol{k}}^{A}\right)^{\dagger}\boldsymbol{\sigma}\Phi_{7,\boldsymbol{k}^{\prime}}^{A}+\nonumber \\
 & +\frac{\left(A_{n}-B_{n}\right)}{2}\left(\Phi_{6,\boldsymbol{k}}^{A}\right)^{\dagger}\boldsymbol{\sigma}\Phi_{6,\boldsymbol{k}^{\prime}}^{A}
\end{align}
The Kondo couplings are then,
\begin{align}
J_{\left\langle LP\right\rangle }^{\Gamma_{7}} & =\frac{J_{0}}{2}\sum_{m=-\infty}^{\infty}\frac{A_{n}+B_{n}}{1+2m\tilde{\Omega}},\nonumber \\
J_{\left\langle LP\right\rangle }^{\Gamma_{6}} & =\frac{J_{0}}{2}\sum_{m=-\infty}^{\infty}\frac{A_{n}-B_{n}}{1+2m\tilde{\Omega}}.
\end{align}
For odd $n$, $B_{n}=0$ due to the properties
of Bessel functions, and we find the form-factor combinations,

\begin{align}
\left\langle M_{x,y}^{n}\!\left(\boldsymbol{k},\boldsymbol{k}^{\prime}\right)\right\rangle  =A_{n}\!\left(\!\begin{array}{cc}
0 & ck_{x}ck_{x}^{\prime}-ck_{y}ck_{y}^{\prime}\\
ck_{x}ck_{x}^{\prime}-ck_{y}ck_{y}^{\prime} & 0
\end{array}\!\right)\!,\nonumber \\
\left\langle M_{z}^{n}\!\left(\boldsymbol{k},\boldsymbol{k}^{\prime}\right)\right\rangle  =A_{n}\!\left(\!\begin{array}{cc}
-ck_{x}ck_{x}^{\prime}-ck_{y}ck_{y}^{\prime} & 0\\
0 & ck_{x}ck_{x}^{\prime}+ck_{y}ck_{y}^{\prime}
\end{array}\!\right)\!,
\end{align}
where here we introduce $ck_i\equiv\cos k_i$.  These terms can be re-decomposed in terms of the $\Gamma_{6,7B}$ form factors.
Therefore, for odd $n$,
\begin{align}
 & \left\langle \boldsymbol{M}^{n\,\text{odd}}\left(\boldsymbol{k},\boldsymbol{k}^{\prime}\right)\right\rangle =\nonumber \\
 & \begin{cases}
\frac{A_{n}}{2}\left[\left(\Phi_{6,\boldsymbol{k}}^{B}\right)^{\dagger}\sigma_{i}\Phi_{7,\boldsymbol{k}^{\prime}}^{B}+\left(\Phi_{7,\boldsymbol{k}}^{B}\right)^{\dagger}\sigma_{i}\Phi_{6,\boldsymbol{k}^{\prime}}^{B}\right] & ,i=x,y\\
\frac{A_{n}}{2}\left[\left(\Phi_{6,\boldsymbol{k}}^{B}\right)^{\dagger}\sigma_{i}\Phi_{6,\boldsymbol{k}^{\prime}}^{B}+\left(\Phi_{7,\boldsymbol{k}}^{B}\right)^{\dagger}\sigma_{i}\Phi_{7,\boldsymbol{k}^{\prime}}^{B}\right] & ,i=z
\end{cases}
\end{align}
and the overall Kondo coupling is,

\begin{equation}
J_{\text{cross}}=\frac{J_{0}}{2}\sum_{m=-\infty}^{\infty}\frac{A_{n}}{1+\left(2m+1\right)\tilde{\Omega}}.
\end{equation}

\begin{figure}
\includegraphics[width=1\columnwidth]{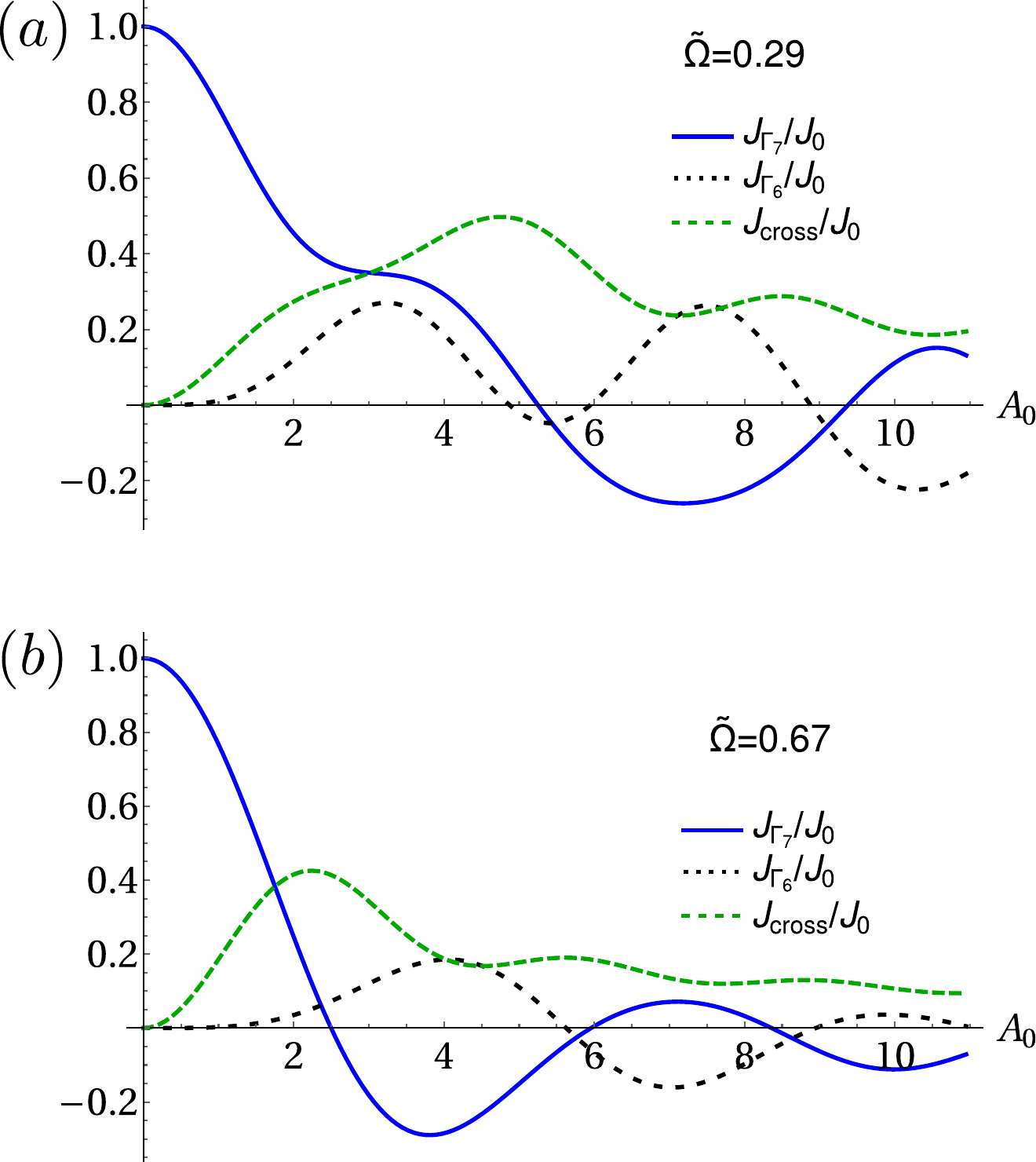}
\caption{The relative strength of the Kondo couplings as function of $A_{0}$
for the Ce model after averaging over LP light for several choices of $\tilde{\Omega}=\Omega/\left|\epsilon_{f}\right|$. \label{fig:Ce-LP-avg}}
\end{figure}

We can now compare the two polarization averaging techniques (LCP/RCP and over LP). The channels are the same in both cases, but the way the $C_{4v}$ symmetry is restored is different. For the LCP/RCP average, the four previously distinct Floquet sectors merge into three, a process that eliminates the chirality inherited from the circular light. For the LP average, a given Floquet sector of fixed $n$ generates more than one channel, separated with the relative strength given by $A_{n}$ and $B_{n}$. Most importantly, the relative strength of the channels depends considerably on the polarization averaging protocol, see Figs.~\ref{fig:Ce-CP-avg} and \ref{fig:Ce-LP-avg}. 

\end{document}